# Electrokinetic oscillatory flow and energy conversion of viscoelastic fluids in microchannels: a linear analysis

Zhaodong Ding[1] and Yongjun Jian[1],†

[1] School of Mathematical Science, Inner Mongolia University, Hohhot, Inner Mongolia 010021, China

We study the electrokinetic flow of viscoelastic fluids subjected to an oscillatory pressure gradient, and particularly focus on the resonance behaviors in the flow. The governing equations are restricted to linear regime so that the velocity and streaming potential fields can be solved analytically. Based on the interaction of viscoelastic shear waves, we explain the mechanism of resonance, and derive a critical Deborah number $De_c = 1/4$ which dictates the occurrence of resonance. Using the Maxwell fluid model, we show that the resonance enhances electrokinetic effects and results in a dramatic increase of electrokinetic energy conversion efficiency. However, by applying the Oldroyd-B fluid model it reveals that the amplification of efficiency is suppressed even for a very small Newtonian solvent contribution. This may be one of the reasons that experimental verification regarding the high efficiency predicted by Bandopadhyay & Chakraborty (*Appl. Phys. Lett.*, vol. 101, 2012, 043905) is unavailable in the literature. Furthermore, the damping effect of solvent viscosity is more significant for higher-order resonances. Introducing the factor of multiple relaxation times, we show that the occurrence of resonances for the streaming potential field and the flow rate are still dominated by $De_c$. For the efficiency in the multi-mode case, the occurrence of resonance is dominated by the Deborah number $De$ and the mode number $N$, and the resonance disappears for small $De$ or large $N$. In addition, a new type of scaling relation between the streaming potential field and EDL thickness can be identified at large $De$.

## 1. Introduction

The transport processes where a charged mobile layer interacts or induces electric fields due to relative motion between the substrate and solution are collectively termed electrokinetics (Hunter 2000). In recent decades, electrokinetic phenomena, including electro-osmosis, streaming potential, electrophoresis, and sedimentation potential, have attracted wide attention and provided a lot of applications in micro- and nanochannels (Schoch *et al.* 2008; Sparreboom *et al.* 2010; Guan *et al.* 2014). The presence of a charged layer close to solids is pivotal to all the electrokinetic phenomena. Most solids spontaneously acquire surface electric charges when brought in contact with a polar medium (Li 2004; Masliyah & Bhattacharjee 2006). These charges are screened by the accumulation of counter-ions near the surface, forming a Stern and diffuse layer, which constitute the so-called electrical double layer (EDL). The characteristic thickness of EDL is commonly known as the Debye length (Jian *et al.* 2017). Charged walls of a channel become increasingly important in micro/nano-scales because of the larger surface to volume ratio (Bocquet & Charlaix 2010). When a liquid is forced to flow through a channel by applying a pressure difference, there is a net flux of ions downstream in the EDL adjacent to the channel walls, which is known as streaming current or advection current. The accumulation of charges at the channel downstream creates an electrical potential difference between two ends of the channel, termed streaming potential that depends on surface charge, electrolyte concentration, and channel dimension (Das *et al*. 2013). This streaming potential field, in turn, generates a current (known as conduction current) to flow back against the direction of the pressure-driven flow. The total current within the cell is only zero if there is no external path for the current, which is often referred to as electroneutrality of current (Goswami & Chakraborty 2010).

† Email address for correspondence: jianyj@imu.edu.cn

Streaming potential phenomena are relevant to suspension rheology (Russel 1978; Sherwood 1980), geophysical two-phase flows through fine porous media (Boléve *et al*. 2007; Sherwood 2007, 2008, 2009; Lac & Sherwood 2009) and the accurate measurement of zeta potentials (Lyklema 1995). Furthermore, streaming potential phenomena provide a mechanism for converting the mechanical energy into electrical energy. Such a mechanism is known as electrokinetic energy conversion (EKEC). The idea of harvesting electrical energy from a fluidic system is not new (Osterle 1964), but it has received considerable interest in recent times owing to the rapid advancements in micro- and nanofluidics (Daiguji *et al.* 2004; Xuan & Li 2006; van der Heyden *et al.* 2006, 2007; Wang & Kang 2010; Chang & Yang 2011; Gillespie 2012; Siria *et al.* 2013). Especially, nanoscale fluidic devices allow the probing of the regime of EDL overlap, where the EKEC efficiency is expected to be highest (Pennathur *et al.* 2007). Although the energy harnessed from a single nanochannel may be rather small, this effect can be substantially magnified if arrays of nanopores or macroscopic nanoporous materials are employed (Yang *et al.* 2003; Daiguji *et al.* 2006).

At present, the EKEC efficiency measured by experiments is relatively low. For example, the highest efficiency of ~3% was obtained by van der Heyden *et al.* (2007) for a 75 nm high channel with dilute KCl aqueous solution under steady pressure-driven flow. In order to improve the EKEC efficiency, a lot of work has been done by employing wall slip (Davidson & Xuan 2008*a*; Ren & Stein 2008), soft nanochannels (Chanda *et al.* 2014; Patwary *et al.* 2016), steric effect (Bandopadhyay & Chakraborty 2011), time-periodic pressure (Goswami & Chakraborty 2010), polymer addition (Nguyen *et al.* 2013), buffer anions effect (Mei *et al.* 2017), transverse magnetic fields (Munshi *et al.* 2009), and layering of large ions near the wall/liquid interface (Gillespie 2012). Recently, Bandopadhyay & Chakraborty (2012) reported a mechanism of massive augmentations in energy harvesting capabilities of nanofluidic devices, through the combined deployment of viscoelastic fluids and oscillatory driving pressure forces. They found that when the forcing frequency of a pressure-driven flow matches with the inverses of the relaxation time scale of a typical viscoelastic fluid, the EKEC efficiency may get giantly amplified due to the complex interplay between the fluid rheology and ionic transport within the EDL, which can be attributed to the viscoelastic resonance phenomenon. Specifically, they predicted that for a slit-type microchannel ($\kappa h$ = 500, with the half channel height $h$ and the Debye length $1/\kappa$), the conversion efficiency can be in the order of 10%, and that for a nanochannel ($\kappa h = 10$) without taking into account surface conductance, the conversion efficiency can be even larger than 95%. This result is controversial, and the mechanism of enhancement needs to be further explored. On the one hand, as remarked by Nguyen *et al.* (2017) the efficiency defined by Bandopadhyay & Chakraborty is thermodynamic efficiency, where no actual power is delivered by the system. The maximal efficiency under the condition of maximal output power at a load resistor is more relevant, and it is in the tune of 24% for a nanochannel (Nguyen *et al.* 2017). Within the linear response regime of electrokinetic flows, it can be proved that this maximum efficiency does not exceed 50% based on thermodynamic analysis (Xuan & Li 2006; Ding *et al.* 2019). On the other hand, more fundamentally, the model used by Bandopadhyay & Chakraborty is a simple Maxwell one that is a considerable simplification of viscoelastic fluid and only valid at low shear rates and frequencies (Castrejón-Pita *et al.* 2003). Some significant deviations from Maxwell behavior are observed at intermediate and high frequency regimes for wormlike micellar solutions, due to Rouse-like behavior of the individual entangled segments (Yesilata *et al.* 2006) and an additional solvent stress (Casanellas & Ortín 2012).

In order to clarify the fundamental aspects of resonance phenomena in electrokinetic oscillatory flows of viscoelastic fluids, we use a generic constitutive model and provide a detailed analysis in the linear regime. The viscoelastic fluid used in practice is usually a suspension formed by various polymers or a solution with fine additives. The macroscopic description for viscoelastic fluids combines the elastic behavior, represented by an elastic spring, with the viscous behavior represented by a dissipative dashpot. Depending on the precise configuration of these primary elements one can obtain different constitutive equations (e.g., Maxwell model, Oldroyd-B model, Giesekus model, Phan-Thien-Tanner model, Johnson-Segalman model, etc). For polymer aqueous solutions, the effect of solvent composition can be investigated by the Oldroyd-B model, which represents a solution of a Maxwellian viscoelastic fluid with a single relaxation time $\lambda$, solved in a Newtonian solvent with constant viscosity (Bird *et al.* 1987).

The laminar oscillatory flows of Maxwell and Oldroyd-B viscoelastic fluids have been studied by Casanellas & Ortín (2011), and exhibit many interesting features that are absent in corresponding flows of Newtonian fluids. Barnes *et al.* (1969) investigated the effect of an oscillatory pressure gradient around a nonzero mean in straight cylinders. The experimental results obtained for dilute aqueous solutions of polyacrylamide showed a dramatic mean flow rate enhancement at particular values of the mean pressure gradient, as a result of 'resonance' effect between the fluid elasticity and the oscillatory driving. Their observations were in qualitative agreement with their own theoretical predictions based on a power series expansion of the velocity and shear rate of a fluid characterized by an apparent viscosity. The mechanisms for flow enhancement, using different fluid models, were subsequently studied by several authors such as Barnes *et al.* (1971), Davies *et al.* (1978), Manero & Walters (1980), Phan-Thien and coworkers (Phan-Thien 1978, 1980, 1981; Phan-Thien & Dudek 1982*a*, *b*), Signer (1991), Andrienko *et al.* (2000), and Herrera (2010).

It is well known that steady and transient flow of most polymeric solutions and melts cannot be adequately described by single mode differential constitutive equations (Bird *et al.* 1987). Hence, multi-mode constitutive equations need to be utilized in order to obtain a better description of real fluids, which possess spectrum of relaxation times rather than a single characteristic time scale. The concept has been around for more than a century in the form of the generalized Maxwell model for linear viscoelasticity (Sadek & Pinho 2019). In particular, the Rouse-like behavior for dilute polymer solutions, as previously mentioned, can be expressed by the generalized Maxwell model. Prost-Domasky & Khomami (1996) analyzed the start up of plane Couette flow and large amplitude oscillatory shear flow of single and multi-mode Maxwell fluids as well as Oldroyd-B fluids by analytical or semi-analytical procedures. Andrienko *et al.* (2000) studied the resonance behavior of a fluid described by upper-convected Maxwell (UCM) model with a discrete spectrum of relaxation times. Moyers-Gonzalez *et al.* (2008, 2009) proposed a nonhomogeneous hemorheological model for human blood, where the effect of multiple relaxation times has been introduced. Recently, Sadek & Pinho (2019) presented an analytical solution for electro-osmotic oscillatory flow of multi-mode upper-convected Maxwell (UCM) fluids in small amplitude oscillatory shear.

In the present work, we investigate the streaming potential and EKEC of viscoelastic fluids by introducing a multi-mode model. The strain-stress relationship takes the following form

$$\left.\begin{array}{l} \boldsymbol{T} = \boldsymbol{T}_s + \boldsymbol{T}_p, \\ \boldsymbol{T}_s = \eta_s \dot{\boldsymbol{\gamma}}, \boldsymbol{T}_p = \sum_{k=1}^{N} \boldsymbol{T}_k, \\ \boldsymbol{T}_k + \lambda_k \stackrel{\nabla}{\boldsymbol{T}_k} = \eta_k \dot{\boldsymbol{\gamma}}. \end{array}\right\} \quad (1.1)$$

Here, $\boldsymbol{T}$ denotes the Cauchy stress tensor, which is decomposed into a polymer contribution $\boldsymbol{T}_p$ and a Newtonian solvent contribution $\boldsymbol{T}_s$. $\dot{\boldsymbol{\gamma}} = \nabla \boldsymbol{V} + (\nabla \boldsymbol{V})^{\dagger}$ is the rate-of-strain tensor with the fluid velocity $\boldsymbol{V}$. The superscript $\nabla$ stands for the upper-convected derivative

$$\stackrel{\nabla}{\boldsymbol{T}} = \frac{\partial \boldsymbol{T}}{\partial t} + \boldsymbol{V} \cdot \nabla \boldsymbol{T} - (\nabla \boldsymbol{V})^{\dagger} \cdot \boldsymbol{T} - \boldsymbol{T} \cdot \nabla \boldsymbol{V}. \quad (1.2)$$

$N$, $\lambda_k$ and $\eta_k$ ($k = 1, \ldots, N$) refer to the number of relaxation times in the spectrum, the $k$th relaxation time and the $k$th partial viscosity, respectively. $\eta_s$ is the constant viscosity of Newtonian solvent. The viscosity of polymer can be expressed in terms of the partial viscosity as $\eta_p = \sum \eta_k$, and the total viscosity of the solution is given by the sum of the solvent and polymer contributions, $\eta = \eta_s + \eta_p$. The solvent viscosity fraction $\eta_s/\eta$ is denoted by $X$, which also known as dimensionless retardation time (Casanellas & Ortín 2011).

Essentially, this model is a multi-mode formulations of UCM model. An important feature of this model is that it allows for a spectrum of relaxation times and contains the Newtonian solvent stress. In the case of $N = 1$ (i.e., single relaxation time), one obtains the Oldroyd-B model. Furthermore, for $N = 1$ and $X = 0$, the Newtonian solvent contribution vanishes, and it is simplified to the UCM model. The other limiting behavior is attained for $X = 1$, where the elastic contribution of the polymer vanishes and the Newtonian relation is recovered. For unidirectional flows, the presented model has a linear strain-stress equation, and fails in predicting nonlinear features (e.g., shear-thinning behavior) that appear at considerably large shear rates in viscoelastic fluids. However, it is a good approximation of viscoelastic fluids at small shear rates, and capable to capture the resonance behavior of viscoelastic fluids, which is of interest to us.

Two relevant dimensionless numbers in viscoelastic flows are the Deborah (*De*) and the Weissenberg (*Wi*) numbers (Morozov & van Saarloos 2007). The Deborah number that sets the interplay between the characteristic relaxation time of fluid and the viscous diffusion time, can be defined as

$$De = \lambda_{\max}\eta/\rho h^2 = \lambda_{\max}/t_v, \qquad (1.3)$$

where $\lambda_{\max}$ represents the largest relaxation time, $\rho$ is the fluid density, $h$ is the channel dimension and $t_v = \rho h^2/\eta$ is a characteristic time of purely viscous effects. Note that the above definition of *De* generalizes the one used by del Río *et al*. (1998) to the case of multiple relaxation times. In the limit of *De* → 0 the fluid relaxes much faster than the typical time scale of the flow and Newtonian flow behavior is recovered. For *De* > 1 the relaxation time of the fluid is larger than the time scale of the flow and fluid elasticity dominates the flow behavior. In shear flows, viscoelastic fluids are also subject to shear-driven normal stresses. The ratio between the normal and shear stresses quantifies the non-linear response of the viscoelastic fluid and is proportional to the Weissenberg number (Galindo-Rosales *et al*. 2014) which can be defined as

$$Wi = \lambda_{\max}\dot{\gamma}_{\max}, \qquad (1.4)$$

based on the largest relaxation time and the maximum local shear rate in this paper. Analogously to the Reynolds (*Re*) number in Newtonian fluids, as increased *Wi* different flow regimes can be explored for viscoelastic fluids (Morozov and van Saarloos, 2007). For *Wi* < 1 the fluid is in the laminar shear flows and the fluid response to external forces is expected to be linear, while for *Wi* > 1 nonlinearities start to become manifest. In the high *Wi* regime elastic instabilities and secondary flows are likely to appear (Torralba *et al*. 2007).

In this paper, we focus on the electrokinetic phenomena induced by an oscillatory flow based on the viscoelastic model including the effects of Newtonian solvent and multiple relaxation times. Some restrictions, e.g., low *Wi* and surface potential, have been proposed so that the fluid response to the pressure gradient and streaming potential is in the linear regime, which makes it relatively easy to solve the governing equations, and thereby allowing a detailed mathematical analysis of the resonance behavior. The remainder of this paper is organized as follows. In §2 the problem formulation for the current study and the analytical solutions of fluid velocity and EDL potential are presented. The expressions of the streaming potential and EKEC efficiency are provided and analyzed further. The elastic resonance behaviors of related physical quantities are shown and characterized in detail in §3. The effects of solvent viscosity and multiple relaxation times on the resonance behaviors are investigated in §3.1 and §3.2, respectively. The validity of the linear analysis is examined in §4. Finally, summary and conclusions of the present investigation are given in §5.

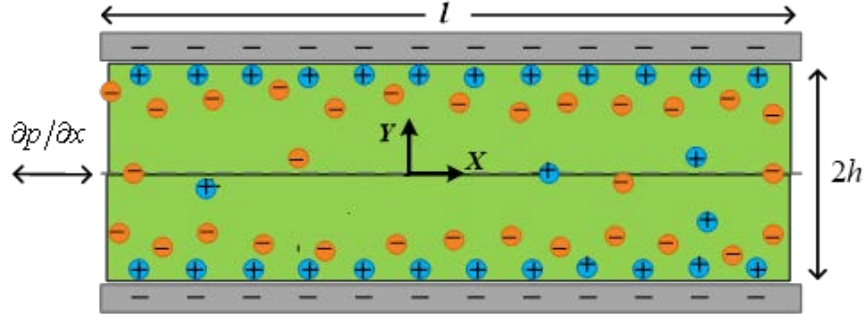

FIGURE 1. (Colour online) Schematic diagram in a microchannel filled with a charged fluid. The EDL forms at the interior of negatively charged walls. A periodic pressure gradient is applied along the channel in the x-direction.

## 2. Theoretical formulation

We consider the electrokinetic transport between two infinitely large parallel plates with a height of $2h$. The fluid medium is viscoelastic one modeling by (1.1), and driven by a harmonically oscillating pressure gradient applied in the direction of the sidewalls (figure 1). A Cartesian system ($x$, $y$, $z$) is constructed, where the $x$-$z$ plane is parallel to the plate surface with the origin at the middle of the microchannel, the $x$ coordinate axis is consistent with the direction of pressure gradient (axial direction), and the $y$ coordinate axis is perpendicular to the plates (transverse direction). The flow is governed by the Navier–Stokes equation with an additional electrical body force and the continuity equation for an incompressible fluid:

$$\left. \begin{array}{l} \rho \dfrac{\partial V}{\partial t} + \rho (V \cdot \nabla) V = -\nabla p + \nabla \cdot T + F_{EK}, \\ \nabla \cdot V = 0. \end{array} \right\} \quad (2.1)$$

$V = (u, v, w)$ denotes the fluid velocity vector. $F_{EK}$ is the electrokinetic body force and is given by (Karniadakis et al. 2005)

$$F_{EK} = \rho_e E - \frac{1}{2} E \cdot E \nabla \varepsilon + \frac{1}{2} \nabla \left( \rho \frac{\partial \varepsilon}{\partial \rho} E \cdot E \right), \quad (2.2)$$

where $\varepsilon$ is the permittivity of the medium, $E$ represents the electric field relating to the electric potential $\phi$ by $E = -\nabla \phi$, $\rho_e = e \sum_i z_i n_i$ is the local net charge density, $e$ is the electronic charge, $n_i$ and $z_i$ are the number density and valence of ion $i$, respectively. The consideration of constant permittivity of the medium reduces the above body force to $F_{EK} = \rho_e E$.

When a fluid contains charged ionic species, we are interested in the movement or mass transfer of the anions and the cations in the fluid flow. In general, the ionic flux satisfies the advection-diffusion-migration equation (i.e. the Nernst–Planck equation)

$$\mathbf{j}_i = n_i V - D_i \nabla n_i - \frac{z_i e n_i D_i}{k_B T} \nabla \phi . \quad (2.3)$$

Here $D_i$ is the diffusivity of ions in the solution, $k_B$ is the Boltzmann constant, $T$ is the absolute temperature, and $\phi$ represents the total electric potential. The first term on the right-hand side of equation (2.3) is the flux due to bulk convection, the second term is due to the concentration gradient (i.e. diffusional process), and the last term is due to migration.

## 2.1. *Electric potential distribution*

The electric potential at a location $(x, y)$ in the channel, given by $\phi(x, y, t)$, arises due to the superposition of the induced external electric potential (streaming potential) and the potential due to the surface charges of the microchannel walls (EDL potential). Assuming that the EDL potential is independent of axial position in the microchannel (which is valid for long microchannels, neglecting any end effects), the total electric potential can be written as

$$\phi(x, y, t) = \psi(y) + [\phi_0 - xE_s(t)] \tag{2.4}$$

where $\psi(y)$ is the EDL potential due to the electric double layer at the equilibrium state corresponding to no fluid motion and no external electric field, $\phi_s \equiv \phi_0 - xE_s(t)$ is the streaming electric potential at a given axial location due to the electric field strength $E_s$ in the absence of the EDLs, $\phi_0$ is the reference value of the potential at $x = 0$, and $E_s$ is the streaming potential field being independent of the position but depending on time since $E_s$ is closely related to the fluid velocity that is time-dependent (see §2.2 and §2.3). Note that (2.4) is an analogue of equation (8.2) of Masliyah & Bhattacharjee (2006) where $E_s$ is independent of time.

The local net charge density $\rho_e$ can be related to the electric potential $\phi$ through the Poisson equation as

$$\nabla^2 \phi = -\frac{\rho_e}{\varepsilon}. \tag{2.5}$$

Introducing the expression (2.4) into the Poisson equation, we obtain

$$\frac{d^2\psi}{dy^2} = -\frac{\rho_e}{\varepsilon}, \tag{2.6}$$

for the slit microchannel.

Considering the zero flux of ions in the *y*-direction at equilibrium, equation (2.3) gives

$$j_{i,y} = n_i v - D_i \frac{dn_i}{dy} - \frac{z_i e n_i D_i}{k_B T} \frac{d\psi}{dy} = 0. \tag{2.7}$$

Furthermore, due to $v = 0$ for laminar parallel flows, one can derive the Boltzmann distribution for the ionic number concentration near a charged surface from (2.7), i.e.

$$n_i = n_{i\infty} \exp\left(-\frac{z_i e \psi}{k_B T}\right) \tag{2.8}$$

with $n_{i\infty}$ denoting the bulk number density of ion $i$ at the neutral state where $\psi = 0$. Combining (2.6) and (2.8), we get the Poisson–Boltzmann equation

$$\frac{d^2\psi}{dy^2} = -\frac{e}{\varepsilon} \sum_i z_i n_{i\infty} \exp\left(-\frac{z_i e \psi}{k_B T}\right) \tag{2.9}$$

for the slit microchannel.

Since the full set of electrokinetic equations are difficult to be solved analytically, three main approximation methodologies have been developed over the years, which are appropriate to low surface electric potential, weak field or flow, and thin double layers respectively. For example, the thin double-layer limit has been applied by Yariv and coworkers (Yariv *et al.* 2011; Schnitzer *et al.* 2012; Schnitzer & Yariv 2016) for streaming potential phenomena. But this approximation cannot be applied in nanochannels where the EDL thickness is comparable to the size of the channel. In our situation, we used the low surface potential assumption, i.e. $|z_i e\psi/k_B T| < 1$, so that the Debye–Hückel approximation can be applied to (2.9), giving (see Masliyah & Bhattacharjee 2006)

$$\frac{d^2\psi}{dy^2} = \kappa^2 \psi \qquad (2.10a)$$

with

$$\kappa^2 = \frac{e^2}{\varepsilon k_B T} \sum_i n_{i\infty} z_i^2 . \qquad (2.10b)$$

The Debye length is $1/\kappa$, characterizing the EDL thickness (typically $\approx$ 10 nm). Equation (2.10a) is the linearized version of the Poisson–Boltzmann equation and gives a satisfactory EDL potential profile (even when linearization may not be justifiable!).

Within the above basic geometry, the EDL potential varies only along the $y$-axis, and the appropriate boundary conditions are

$$\psi = \zeta, \text{ at } y = \pm h. \qquad (2.10c)$$

$\zeta$ is the zeta potential at the walls. Thus, the solution to (2.10a) subject to the conditions in (2.10c) is

$$\psi(y) = \zeta \frac{\cosh(\kappa y)}{\cosh(\kappa h)}, \quad -h \leq y \leq h. \qquad (2.11)$$

For the determination of the streaming potential field in (2.4), the velocity field and the condition of electroneutrality of current are required. The expression of the streaming potential field and its analysis will be provided in §2.3.

## 2.2. *Fluid velocity and flow rate*

For the velocity field, the laminar regime is achieved by applying small driving ($Wi < 1$), which corresponds to small amplitudes of the imposed pressure gradient (Casanellas & Ortín 2011; Casanellas & Ortín 2012). Thus the flow is parallel to the walls and along the $x$-direction. The velocity can be assumed as $\mathbf{V} = (u(y, t), 0, 0)$. The continuity equation is satisfied automatically. The strain-stress relationship (1.1) of viscoelastic fluid is linear for this unidirectional flow, and substituting it into the Navier–Stokes equation, we obtain

$$\left. \begin{array}{l} \rho \dfrac{\partial u}{\partial t} = -\dfrac{\partial p}{\partial x} + \eta_s \dfrac{\partial^2 u}{\partial y^2} + \sum_{k=1}^{N} \dfrac{\partial \tau_k}{\partial y} + \rho_e E_s, \\[6pt] \left(1 + \lambda_k \dfrac{\partial}{\partial t}\right)\tau_k = \eta_k \dfrac{\partial u}{\partial y}, \end{array} \right\} \qquad (2.12a)$$

in which $\tau_k$ is the $xy$ component of $\mathbf{T}_k$, and $E_s = -\partial \phi_s/\partial x$ is the streaming potential field along the $x$-direction and considered to be uniform in the channel. Assume a harmonic pressure gradient $\partial p/\partial x = \cos(\omega t)\, dp_0/dx = \Re\big(dp_0/dx \exp(i\omega t)\big)$ using complex variables, where $dp_0/dx$ is the amplitude of applied pressure gradient and $\omega$ is the frequency of oscillation. The velocity field, stress tensor and streaming potential field can be written in complex forms as

$$u(y,t) = \Re\big(u_0(y)e^{i\omega t}\big), \; \tau_k(y,t) = \Re\big(\tau_k^0(y)e^{i\omega t}\big), \; E_s(t) = \Re\big(E_0 e^{i\omega t}\big), \qquad (2.12b)$$

where $\Re$ denotes the real part of a complex number. Substituting these expressions in (2.12a) leads to

$$i\rho\omega u_0 = -\frac{\mathrm{d}p_0}{\mathrm{d}x} + \eta_s \frac{\mathrm{d}^2 u_0}{\mathrm{d}y^2} + \sum_{k=1}^{N} \frac{\mathrm{d}\tau_k^0}{\mathrm{d}y} + \rho_e E_0,$$
$$(1+i\omega\lambda_k)\tau_k^0 = \eta_k \frac{\mathrm{d}u_0}{\mathrm{d}y},$$
(2.13)

The subsequent analysis will be made in the frequency domain.

Introducing the following dimensionless groups

$$Y = \frac{y}{h}, \quad \Psi = \frac{\psi}{\zeta}, \quad X = \frac{\eta_s}{\eta}, \quad \xi_k = \frac{\eta_k}{\eta}, \quad \bar{\omega} = \omega\lambda_{\max}, \quad K = \kappa h,$$
$$\bar{E}_0 = \frac{E_0}{E_m}, \quad \bar{u}_0 = \frac{u_0}{u_m}, \quad \text{with } u_m = -\frac{h^2}{\eta}\frac{\mathrm{d}p_0}{\mathrm{d}x} \text{ and } E_m = -\frac{h^2}{\varepsilon\zeta}\frac{\mathrm{d}p_0}{\mathrm{d}x},$$
(2.14)

and eliminating $\tau_k^0$ from (2.13) yield

$$\frac{\mathrm{d}^2 \bar{u}_0}{\mathrm{d}Y^2} - \frac{i\bar{\omega}}{De}\left(X + \sum_{k=1}^{N}\frac{\xi_k}{1+i\omega\lambda_k}\right)^{-1}\bar{u}_0 = \left(X + \sum_{k=1}^{N}\frac{\xi_k}{1+i\omega\lambda_k}\right)^{-1}\left[-1 + K^2 \bar{E}_0 \frac{\cosh(KY)}{\cosh(K)}\right]. \quad (2.15)$$

Here $\rho_e = -\varepsilon\kappa^2\psi$ has been used. $\bar{\omega}$ is the non-dimensional frequency, $K$ denotes inverse of the normalized EDL thickness, and $De$ is the Deborah number defined in (1.3). Furthermore, through defining the following dimensionless parameter

$$\alpha = \sqrt{-\frac{i\bar{\omega}}{De}}\left(X + \sum_{k=1}^{N}\frac{\xi_k}{1+i\omega\lambda_k}\right)^{-1/2}, \quad (2.16)$$

the solution of (2.15) subject to the no-slip conditions at walls can be expressed as

$$\bar{u}_0(Y) = \frac{De}{i\bar{\omega}}\left[1 - \frac{\cos(\alpha Y)}{\cos(\alpha)}\right] - \frac{\alpha^2 De K^2 \bar{E}_0}{i\bar{\omega}(K^2+\alpha^2)}\left[\frac{\cosh(KY)}{\cosh(K)} - \frac{\cos(\alpha Y)}{\cos(\alpha)}\right] \triangleq \bar{u}_{P0}(Y) + \bar{u}_{E0}(Y)\cdot\bar{E}_0. \quad (2.17)$$

Equation (2.17) gives the complex amplitude of the axial velocity. The first term is the liquid velocity due to the imposed pressure gradient. The second term represents an additional convective transport of the fluid medium due to the induced streaming potential field. Then, the complex amplitude of dimensionless flow rate (per unit width of channel) scaled by $hu_m$ is given by

$$\bar{Q}_0 = 2\int_0^1 \bar{u}_0(Y)\mathrm{d}Y = \frac{2De}{i\bar{\omega}}\left[1 - \frac{\tan(\alpha)}{\alpha}\right] - \frac{2\alpha^2 De K^2 \bar{E}_0}{i\bar{\omega}(K^2+\alpha^2)}\left[\frac{\tanh(K)}{K} - \frac{\tan(\alpha)}{\alpha}\right] \triangleq \bar{Q}_{P0} + \bar{Q}_{E0}. \quad (2.18)$$

2.3. *Streaming potential field and scaling laws*

As stated in the introduction, when the flow is driven by a pressure gradient, the streaming current and streaming potential can be established and pressure-to-voltage energy conversion can be achieved. Using (2.3), the current density in the $x$-direction can be expressed as

$$i_x = e\sum_i z_i j_{i,x} + \frac{\sigma_{\text{Stern}}}{h}E_s = eu\sum_i z_i n_i - e\sum_i D_i z_i \frac{\partial n_i}{\partial x} + \frac{e^2}{k_B T}E_s\sum_i z_i^2 D_i n_i + \frac{\sigma_{\text{Stern}}}{h}E_s \quad (2.19)$$

where an additional term (i.e. the last one), accounting for the Stern layer conductance with a surface conductivity $\sigma_{\text{Stern}}$, is introduced into the traditional current density equation (see Davidson & Xuan 2008b). The total current per unit width of the slit channel is given by

$$I = 2\int_0^h i_x \, dy. \tag{2.20}$$

For simplicity, we consider a symmetric electrolyte ($z_+ = -z_- = z$) with the same ionic diffusion coefficients $D$. In the present flow system, it is assumed that there is no variation in the ionic concentration along the length of the channel. Consequently, $\partial n_i/\partial x = 0$, and the second term of equation (2.19) drops out. Thus, the current density reduces to

$$\begin{aligned} i_x &= zeu(n_+ - n_-) + \frac{e^2 z^2 D}{k_B T} E_s (n_+ + n_-) + \frac{\sigma_{\text{Stern}}}{h} E_s \\ &= \rho_e u + \sigma_b \cosh\left(\frac{ze\psi}{k_B T}\right) E_s + \frac{\sigma_{\text{Stern}}}{h} E_s, \end{aligned} \tag{2.21a}$$

where the Boltzmann distribution in (2.8) has been used,

$$\sigma = \sigma_b \cosh\left(\frac{ze\psi}{k_B T}\right) \quad \text{and} \quad \sigma_b = \frac{2e^2 z^2 n_\infty D}{k_B T} \tag{2.21b}$$

are the local and bulk conductivity of the fluid, respectively. Note that the factor $\cosh(ze\psi/k_B T)$ reflects the influence of diffuse charge cloud on the bulk conductivity.

Applying the Taylor series for cosh term

$$\cosh\left(\frac{ze\psi}{k_B T}\right) = 1 + \frac{1}{2!}\left(\frac{ze\psi}{k_B T}\right)^2 + O(\psi^4) \tag{2.22}$$

to the first one in (2.21b), one can find that it is $O(\zeta^2)$ for the modification of the bulk conductivity from the diffuse layer, so the effect of the diffuse charges on the bulk conductivity can be ignored and one gets $\sigma \approx \sigma_b$ based on the low surface potential assumption.

To evaluate the relative importance between Stern layer conductance and bulk conductance, we introduce the Dukhin number $Du = \sigma_{\text{Stern}}/h\,\sigma_b$. From (2.21b), we have $Du = \sigma_{\text{Stern}}/h\,\Lambda c_b$ where $c_b$ is the ionic concentration of the bulk fluid and $\Lambda = 2e^2 z^2 N_A D/k_B T$ is the molar conductivity with $N_A$ the Avogadro's number. To examine quantitatively the effect of Stern layer conductance, we consider KCl solution with the concentration of $c_b = 10^{-4}$ M as an example. The ionic diffusion coefficient $D$ is approximately equal to $1.9 \times 10^{-9}$ m$^2$s$^{-1}$ (Daiguji et al. 2004). The molar conductivity can be calculated as $\Lambda = 0.0144$ S·m$^2$·mol$^{-1}$ at temperature $T = 298$ K (see also Davidson & Xuan 2008b). The half height of the channel $h = 1$ μm. The value of $\sigma_{\text{Stern}}$ is in the range of $10-30$ pS in the experiment of van der Heyden et al. (2007). Thus, the Dukhin number is between 0.01 and 0.03. This indicates that the Stern layer conductance may have an important effect only for tiny channel size and extremely low ionic concentration ($< 10^{-4}$ M). Here we ignore the influence of Stern layer conductance.

Integrating the current density over the channel section in (2.20) and applying the complex forms in (2.12b), one can obtain the complex amplitude of total ionic current

$$I_0 = 2\int_0^h \rho_e u_0 \, dy + 2h\sigma_b E_0, \tag{2.23a}$$

from which the amplitudes of streaming current and conduction current through the channel can be identified as

$$I_{s0} = 2\int_0^h \rho_e u_0 \, dy \quad \text{and} \quad I_{c0} = 2h\sigma_b E_0, \tag{2.23b}$$

respectively. By using the condition of electroneutrality of current ($I_{s0} + I_{c0} = 0$), the dimensionless variables (2.14) and the velocity field (2.17), the amplitude of dimensionless streaming potential field can be expressed as

$$\bar{E}_0 = \frac{\int_0^1 \Psi(Y)\bar{u}_{P0}(Y)\,dY}{R - \int_0^1 \Psi(Y)\bar{u}_{E0}(Y)\,dY}, \tag{2.24}$$

where

$$R = \frac{\eta D}{\varepsilon \zeta^2} \tag{2.25}$$

is the inverse ionic Péclet number, characterizing the ratio of the conduction current to the streaming current (Chanda *et al.* 2014; Chakraborty & Das 2008). Furthermore, using the analytical solutions of the velocity and the EDL potential $\Psi$, we get

$$\int_0^1 \Psi(Y)\bar{u}_{P0}(Y)\,dY = \frac{\alpha De}{i\bar{\omega}(K^2 + \alpha^2)}\left[\frac{\alpha}{K}\tanh(K) - \tan(\alpha)\right] \tag{2.26a}$$

and

$$\int_0^1 \Psi(Y)\bar{u}_{E0}(Y)\,dY = -\frac{\alpha^2 K^2 De}{i\bar{\omega}(K^2 + \alpha^2)}\left[\frac{2K + \sinh(2K)}{4K\cosh^2(K)} - \frac{K\sinh(K)\cos(\alpha) + \alpha\cosh(K)\sin(\alpha)}{(K^2 + \alpha^2)\cosh(K)\cos(\alpha)}\right]. \tag{2.26b}$$

For viscous fluids case (e.g., Newtonian fluids), the effect of the induced streaming potential on the velocity is negligible compared with that of pressure gradient (i.e. $\bar{u}_0 \approx \bar{u}_{P0}$) (see figure 5 for low *De*), so the streaming potential field in (2.24) can be approximated as

$$\bar{E}_0 \approx \frac{1}{R}\int_0^1 \Psi(Y)\bar{u}_{P0}(Y)\,dY. \tag{2.27}$$

Furthermore, from the definition of *R*, it can be deduced that $\bar{E}_0$ is proportional to $\zeta^2$. This indicates that the contribution of streaming potential to the velocity is the order of $O(\zeta^2)$ in (2.17). However, if fluid elasticity dominates the flow behavior, resonances may occur and result in huge increase in the amplitudes of flow rate and streaming potential (see figures 6 and 7). At this time, the effect of the streaming potential field on the velocity is significant and cannot be ignored (see figure 5 for high *De*). As a result, the approximation (2.27) and thereby the relationship $\bar{E}_0 \propto R^{-1}$ or $\bar{E}_0 \propto \zeta^2$ will no longer hold.

Figure 2 displays the variations of maximum amplitude of streaming potential field, defined by $\bar{E}_{0,\max} = \max|\bar{E}_0(\omega)|$, as a function of the ionic Péclet number $R^{-1}$ at different Deborah number. Here the maximum amplitude occurs at the first-order resonance frequency (see figures 7*a* and 8*b*). At other frequencies, similar behavior can also be observed. For comparison, both of the formulas (2.24) and (2.27) have been applied in this figure. By using the physical properties: $D = 1.9 \times 10^{-9}$ m²s$^{-1}$, $\eta = 1.0 \times 10^{-3}$ kg/ms, $\varepsilon = 7 \times 10^{-10}$ C/Vm, and $\zeta = 20$ mV, we get $R \approx 7$. To

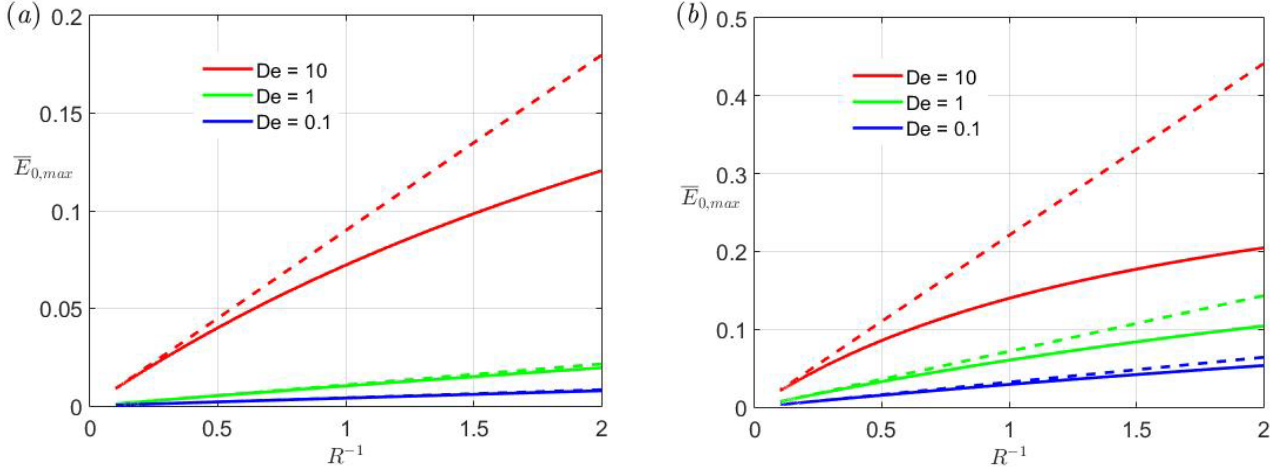

FIGURE 2. (Colour online) Scaling relations between the dimensionless streaming potential field and the ionic Péclet number $R^{-1}$ at different Deborah number, where $\bar{E}_{0,\max} = \max|\bar{E}_0(\omega)|$ with $\bar{E}_0$ provided by (2.24) (solid line) and the approximation (2.27) (dashed line), respectively. (a) $K = 15$, $X = 0$; (b) $K = 5$, $X = 0.1$.

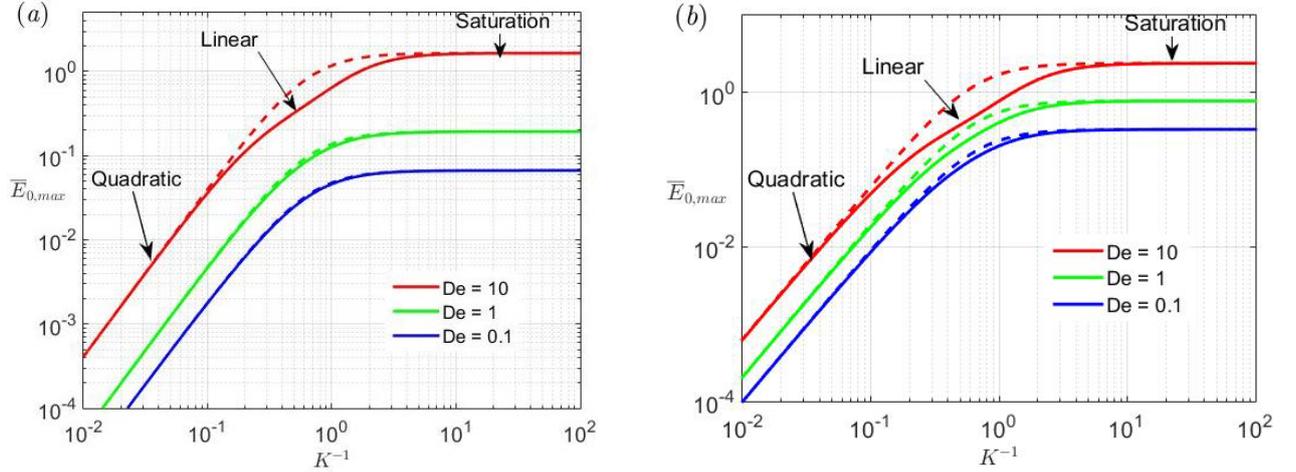

FIGURE 3. (Colour online) Scaling relations between the dimensionless streaming potential field and EDL thickness $K^{-1}$ at different Deborah number, where $\bar{E}_{0,\max} = \max|\bar{E}_0(\omega)|$ with $\bar{E}_0$ provided by (2.24) (solid line) and the approximation (2.27) (dashed line), respectively. (a) $R = 5$, $X = 0$; (b) $R = 1$, $X = 0.1$.

extend the analysis, the range of $R$ is considered to be from 0.5 to 10 (equivalently, $R^{-1}$ is from 0.1 to 2). We observe that when the fluid viscosity dominates the flow behavior (represented by low Deborah number), the variations of streaming potential given by these two formulas are consistent, and a linear relationship $\bar{E}_{0,\max} \propto R^{-1}$ can be invoked, as expected. However, the approximation (2.27) is gradually failing with the increase of *De* due to the occurrence of elastic resonance, and significant deviations from the linear relationship can be observed for streaming potential field (the resonance behavior will be analyzed in detail in §3).

In figure 3 we show the scaling relations between the streaming potential field $\bar{E}_{0,\max}$ and the dimensionless EDL thickness $K^{-1}$ at different Deborah number. For viscous fluids case (low Deborah number), there exist two distinct scaling regimes. For $K^{-1} < 1$, we observe that $\bar{E}_{0,\max}$ increases with $K^{-1}$ with $\bar{E}_{0,\max} \sim K^{-2}$, which was named as the Quadratic Regime by Das *et al.* (2013). While for $K^{-1} > 1$, i.e. large EDL thickness relative to channel size, $\bar{E}_{0,\max}$ saturates and show no further variation with $K^{-1}$. Hence this regime was called the Saturation Regime (Das *et al.*

2013). When the elasticity of fluid is dominant (represented by high Deborah number, e.g., $De = 10$), a new type of regime can be witnessed in addition to the above two regimes, which occurs near $K^{-1} \approx 1$ (the onset of EDLs overlap) and does not appear in the case of viscous fluids (see the dashed line in figure 3). In this third regime, $\bar{E}_{0,\max}$ increases with $K^{-1}$ with $\bar{E}_{0,\max} \sim K^{-1}$. We shall call this regime as the Linear Regime.

### 2.4. *Electrokinetic energy conversion (EKEC)*

Now consider such a fluid system connected to an external load (with resistance $R_L$). Due to the establishment of streaming current and potential, the mechanical energy is converted into the electric energy through this external load. This fluid system can be treated as a battery of an ideal current source plus an internal resistance $R_{ch}$, and an equivalent circuit has been presented (see, e.g., van der Heyden *et al.* 2007). The streaming current ($I_s$) is the one through the circuit when the external resistance $R_L$ is zero, i.e. $I_s$ can be viewed as a short circuit current in the circuit where the voltage $\Delta \phi$ is zero. The streaming potential ($\Delta \phi_s$) is equal to the voltage on $R_L$ when $R_L$ value is infinite, i.e. $\Delta \phi_s$ can be viewed as an open-circuit voltage in the circuit where the current $I$ is zero. For this circuit, the instantaneous output power $W_{out} = I\Delta \phi$ where $I$ is the current through $R_L$ and $\Delta \phi$ is the voltage on $R_L$. It is known that the maximum power transfer occurs at $R_L = R_{ch}$ (Olthuis *et al.* 2005; van der Heyden *et al.* 2007). In that case the current through the external load is half of the streaming current, i.e. $I = I_s/2$, and the voltage on $R_L$ is half of the streaming potential, i.e. $\Delta \phi = \Delta \phi_s/2$. Note that in our system the current and voltage vary with time. The time-averaged maximum output power can be expressed as

$$W_{out,max} = \frac{1}{4}\langle I_s \Delta \phi_s \rangle = \frac{\omega}{8\pi} \int_0^{2\pi/\omega} I_s(t) \Delta \phi_s(t) dt \qquad (2.28)$$

over a cycle. Using the complex forms $I_s(t) = \Re(I_{s0}e^{i\omega t}) = (I_{s0}e^{i\omega t} + I_{s0}^* e^{-i\omega t})/2$, $E_s(t) = \Re(E_0 e^{i\omega t}) = (E_0 e^{i\omega t} + E_0^* e^{-i\omega t})/2$ (* indicating complex conjugate), the expression $\Delta \phi_s(t) = -lE_s(t)$ with length $l$ of the channel, and $I_{s0} = -I_{c0} = -2h\sigma_b E_0$, we obtain

$$W_{out,max} = \frac{1}{4} h l \sigma_b |E_0|^2. \qquad (2.29)$$

With the flow rate $Q = \Re(Q_0 e^{i\omega t})$ through the channel under the pressure difference $\Delta p = l \, \partial p/\partial x = l\Re(dp_0/dx \exp(i\omega t))$, the time-averaged hydrodynamic power is

$$W_{in} = \langle Q(-\Delta p) \rangle = -\frac{l}{2} \frac{dp_0}{dx} \Re(Q_0). \qquad (2.30)$$

Note that the negative sign in the above expression is due to the fact that the direction of fluid flow is opposite to the pressure gradient. Furthermore, using (2.29), (2.30) and the non-dimensional forms (2.14) and (2.18), the maximum power transfer efficiency can be calculated:

$$\xi = \frac{W_{out,max}}{W_{in}} = \frac{RK^2 |\bar{E}_0|^2}{2\Re(\bar{Q}_0)}. \qquad (2.31a)$$

It is worth noting that maximum EKEC efficiency has been defined and calculated in different ways in the literature. For example, Xuan & Li (2006) calculated the maximum conversion efficiency based on thermodynamics while the work of van der Heyden *et al.* (2006) is based on the circuit. A few researchers (e.g., Chanda *et al.* 2014; Buren *et al.* 2018; Koranlou *et al.* 2019) defined the efficiency $\xi$ by purely pressure-driven volume flow rate without the retardation induced by the back electro-osmotic transport.

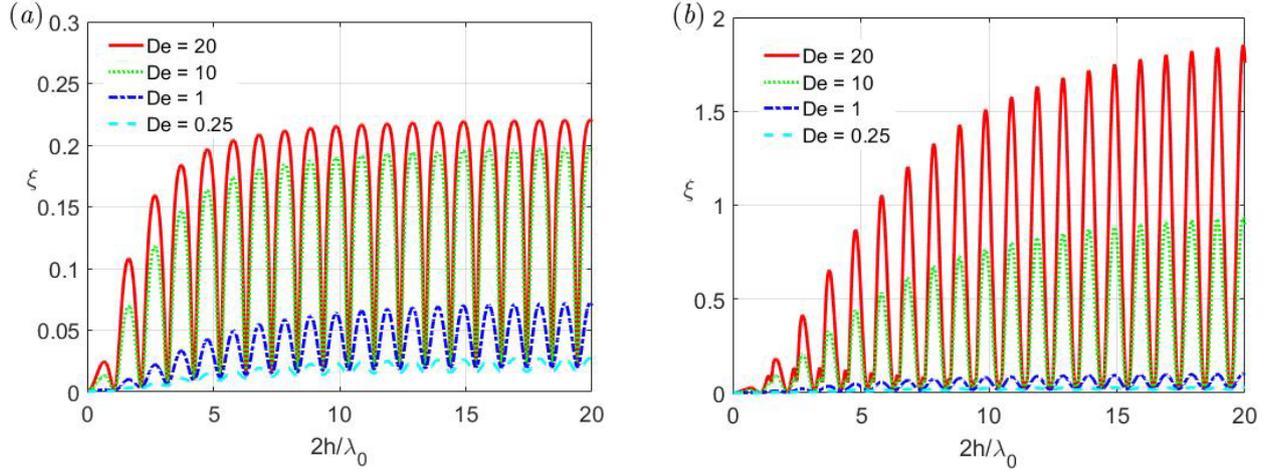

FIGURE 4. (Colour online) EKEC efficiencies calculated by different definitions for Maxwell fluids ($X = 0$) at different $De$ number, where (*a*) the definition (2.31*a*) is used and (*b*) the definition (2.31*b*) with the purely pressure-driven volume flow rate is used. Note that the meaning of the abscissa $2h/\lambda_0$ can be found in §3.

In our recent work (Ding *et al.* 2019), we demonstrated the equivalence of thermodynamic and electric circuit analyses for calculating maximum conversion efficiency in the linear response regime. In fact, the calculation of maximum EKEC efficiency based on the circuit has been adopted in this work. For comparison, we also provide the expression of maximum EKEC efficiency based on the purely pressure-driven flow rate as follows

$$\xi' = \frac{1}{4}\frac{\langle I_s \Delta\phi_s \rangle}{\langle Q_p (-\Delta p) \rangle} = \frac{RK^2 |\bar{E}_0|^2}{2\Re(\bar{Q}_{p0})}, \quad (2.31b)$$

with

$$\bar{Q}_{p0} = \frac{2De}{i\bar{\omega}}\left[1 - \frac{\tan(\alpha)}{\alpha}\right] \quad (2.32)$$

the purely pressure-driven dimensionless flow rate (see also (2.18)).

For viscous fluids, because the effect of induced streaming potential on the velocity is negligible compared with that of pressure gradient, one can get $\bar{Q}_0 \approx \bar{Q}_{p0}$. As a result, the efficiencies calculated by (2.31*a*) and (2.31*b*) are approximately equal (see also Olthuis *et al.* 2005; Berli 2010; Ding *et al.* 2019). However, for viscoelastic fluids that exhibit resonance behavior, the effect of the streaming potential on the velocity cannot be ignored and these two different definitions on the EKEC efficiency will lead to significant differences. In particular, when employing the definition (2.31*b*) to calculate the EKEC efficiency, it predicts (incorrectly) a maximum efficiency above 100% (see figure 4), which violates the law of thermodynamics. This means that the conversion efficiency defined by the purely pressure-driven flow rate does not apply to this situation.

## 3. Resonance phenomenon in electrokinetic transport

It has been predicted by Bandopadhyay & Chakraborty (2012) that the EKEC efficiency may get giantly amplified due to the resonance phenomenon of viscoelastic fluids when subjected to periodic external forcing. However, the model they used is only valid in a narrow range of the parameters (e.g., low shear rates and frequencies), and the effects of resonance on the flow of liquid and streaming potential field in microfluidic channels still remain unclear. By using a viscoelastic fluid model that includes the effects of Newtonian solvent and multiple relaxation times, we aim to address this issue at medium and high frequencies, which is especially important because the resonance phenomena usually occur in this range. It is worth emphasizing, however, that the low shear rate is still assumed in our situation so that the problem is in the linear regime and thus can be treated analytically.

The laminar oscillatory flows of single-mode Maxwell and Oldroyd-B fluids between two infinitely large parallel plates were analyzed by Casanellas & Ortín (2011) from the perspective of viscoelastic shear waves. This perspective brings new physical insight that is useful to understand the mechanism of resonance phenomena. We apply this method to explore the electrokinetic effects in viscoelastic oscillatory flows, and further extend it to the case of multi-mode oscillatory flows. First, we prove that the time-periodic pressure driven electrokinetic flow is equivalent to the flow that results from the synchronous oscillatory sidewalls with the same frequency, observed in the reference frame of the sidewalls. This proof is given in appendix A. Note that this equivalence holds only in the linear regime (see appendix A). Then the flow can be regarded as the result of the interference in time and space of the shear waves generated by the oscillatory sidewalls. The dimensionless parameter $\alpha$ defined in (2.16) plays a key role in the analysis. Obviously, $\alpha$ is a complex number, and can be rewritten as

$$\alpha = \operatorname{Re}(\alpha) + i \cdot \operatorname{Im}(\alpha) = \frac{2\pi h}{\lambda_0} - i\frac{h}{x_0}, \tag{3.1}$$

where $\lambda_0$ and $x_0$ can be identified as the wavelength and damping length of the transverse wave, and given by

$$\frac{\lambda_0}{2\pi} = \sqrt{\frac{2h^2 De}{\bar{\omega}}} \sqrt{\frac{\Omega_1^2 + \Omega_2^2}{\Omega_2 + \sqrt{\Omega_1^2 + \Omega_2^2}}}, \quad x_0 = \sqrt{\frac{2h^2 De}{\bar{\omega}}} \sqrt{\frac{\Omega_1^2 + \Omega_2^2}{-\Omega_2 + \sqrt{\Omega_1^2 + \Omega_2^2}}}, \tag{3.2a,b}$$

with

$$\Omega_1 = X + \sum_{k=1}^{N} \frac{\xi_k}{1 + (\omega \lambda_k)^2}, \quad \Omega_2 = \sum_{k=1}^{N} \frac{\omega \lambda_k \xi_k}{1 + (\omega \lambda_k)^2}. \tag{3.3a,b}$$

The viscosity ratios $X$ and $\xi_k$ satisfy $X + \sum_{k=1}^{N} \xi_k = 1$. Furthermore, we get

$$\left|\frac{\operatorname{Re}(\alpha)}{\operatorname{Im}(\alpha)}\right| = \frac{2\pi x_0}{\lambda_0} = \frac{\Omega_2}{\Omega_1} + \sqrt{1 + \left(\frac{\Omega_2}{\Omega_1}\right)^2}. \tag{3.4}$$

The ratio $2\pi x_0/\lambda_0$ depends on the value of $\Omega_2/\Omega_1$, and is larger than one except for the Newtonian case $X = 1$, where $\xi_k = 0$ and thus $\Omega_2/\Omega_1 = 0$. In particular for Maxwell fluids ($X = 0$, $N = 1$),

$$\left|\frac{\operatorname{Re}(\alpha)}{\operatorname{Im}(\alpha)}\right| = \frac{2\pi x_0}{\lambda_0} = \bar{\omega} + \sqrt{1 + \bar{\omega}^2} \gg 1 \text{ for large } \bar{\omega}. \tag{3.5}$$

The fact that this ratio is larger than one is important and reveals that viscoelastic shear waves are always underdamped. In contrast, transverse oscillations of Newtonian fluids are overdamped, making resonance impossible.

Now, we concern on the situation of resonance. From the expression (2.17) of the velocity, we deduce that the resonance occurs at $\cos(\alpha) = 0$. By means of Euler formula,

$$\cos(\alpha) = \frac{e^{\text{Im}(\alpha)} + e^{-\text{Im}(\alpha)}}{2}\cos(\text{Re}(\alpha)) + i\frac{e^{-\text{Im}(\alpha)} - e^{\text{Im}(\alpha)}}{2}\sin(\text{Re}(\alpha)), \quad (3.6)$$

so

$$\cos(\alpha) = 0 \quad \Leftrightarrow \quad \text{Im}(\alpha) = 0 \text{ and } \text{Re}(\alpha) = n\pi - \frac{\pi}{2}, \quad n = 1, 2, \ldots. \quad (3.7)$$

However, $\text{Im}(\alpha) = -h/x_0 = 0$ is an unrealistic requirement, so we resort to a weak version

$$\cos(\alpha) \approx 0 \quad \Leftrightarrow \quad |\text{Im}(\alpha)| \ll 1 \text{ and } \text{Re}(\alpha) = n\pi - \frac{\pi}{2}, \quad n = 1, 2, \ldots. \quad (3.8)$$

Next, we analyze the meanings of the two conditions in (3.8) based on the viscoelastic shear waves. At this point we introduce a viscoelastic Stokes parameter

$$\Lambda_{ve} = \frac{h}{x_0} = |\text{Im}(\alpha)|, \quad (3.9)$$

as defined in Casanellas & Ortín (2011). The parameter $\Lambda_{ve}$ represents the ratio of the transverse size of systems to the extension of viscoelastic shear waves generated either by the oscillatory pressure gradient or the moving sidewalls. 'Narrow' systems correspond to $\Lambda_{ve} < 1$, where the viscoelastic shear waves extend through the whole system, and 'wide' systems correspond to $\Lambda_{ve} > 1$, where the shear waves are attenuated at the center of the channel. For 'narrow' systems the viscoelastic shear waves generated at both plates superpose themselves and originate an interference pattern inside the fluid domain. This leads to a resonance behavior with a huge increase of the velocity amplitude at particular frequencies (see figure 5). In contrast, for 'wide' systems, the interaction of the shear waves is weak, the viscous behavior is manifested, and no resonance occurs. Thus, one of the conditions in (3.8) for resonance (i.e., $|\text{Im}(\alpha)| < 1$) means a 'narrow' system from the perspective of viscoelastic shear waves. The other condition in (3.8) for resonance reads

$$\frac{2h}{\lambda_0} = n - \frac{1}{2}, \quad n = 1, 2, \ldots \quad (3.10)$$

which means that the size of the channel $2h$ is a half integer multiple of the wavelength $\lambda_0$, and can be viewed as the constructive interference of the shear waves. The number $n$ can be called resonance order. In contrast, the condition

$$\frac{2h}{\lambda_0} = n, \quad n = 1, 2, \ldots \quad (3.11)$$

implies the destructive interference of the shear waves. Further illustrations on the constructive and destructive interferences are given in Appendix A.

From these conditions, one can find readily that the resonance behavior is closely related to the setup dimensions, fluid parameters, and applied driving frequency.

### 3.1. *Single-mode Maxwell and Oldroyd-B fluids cases*

In order to facilitate the analysis of the role of elasticity, we first investigate the resonance phenomenon in electrokinetic transport based on the single-mode Maxwell and Oldroyd-B fluids. Such analyses are useful for guiding the investigation of multi-mode viscoelastic fluids.

In these special cases, the wavelength and damping length of transverse waves reduce to

$$\frac{\lambda_0}{2\pi} = \sqrt{\frac{2h^2 De}{\bar{\omega}}} \sqrt{\frac{1+X^2\bar{\omega}^2}{(1-X)\bar{\omega}+\sqrt{(1+\bar{\omega}^2)(1+X^2\bar{\omega}^2)}}}, \quad (3.12)$$

and

$$x_0 = \sqrt{\frac{2h^2 De}{\bar{\omega}}} \sqrt{\frac{1+X^2\bar{\omega}^2}{(X-1)\bar{\omega}+\sqrt{(1+\bar{\omega}^2)(1+X^2\bar{\omega}^2)}}}, \quad (3.13)$$

which reproduce the results of Casanellas & Ortín (2011) for Oldroyd-B fluids (pay attention to the differences in the definitions and symbols of the dimensionless variables used). The viscoelastic Stokes parameter can be rewritten in terms of the Deborah number $De$, non-dimensional frequency $\bar{\omega}$, and solvent viscosity fraction $X$, as

$$\Lambda_{ve} = \frac{1}{\sqrt{De}} f(\bar{\omega}, X), \quad (3.14)$$

with

$$f(\bar{\omega}, X) = \sqrt{\frac{(X-1)\bar{\omega}^2 + \bar{\omega}\sqrt{(1+\bar{\omega}^2)(1+X^2\bar{\omega}^2)}}{2(1+X^2\bar{\omega}^2)}}. \quad (3.15)$$

Particularly, for Maxwell fluids ($X = 0$), we have

$$|f(\bar{\omega}, X=0)| = \sqrt{\frac{\bar{\omega}\sqrt{(1+\bar{\omega}^2)} - \bar{\omega}^2}{2}} = \sqrt{\frac{\bar{\omega}}{2(\sqrt{(1+\bar{\omega}^2)} + \bar{\omega})}} \leq \frac{1}{2}, \quad (3.16)$$

and

$$f(\bar{\omega}, X=0) \approx \frac{1}{2} \quad \text{as} \quad \bar{\omega} \gg 1. \quad (3.17)$$

It is known that viscoelastic fluids, described by the Maxwell model, have different flow regimes depending on the value of the parameter $De$ (de Haro *et al.* 1996; del Río *et al.* 1998). A critical value $De_c = 1/11.64$, determining whether a dissipative behavior prevails or a resonance at a given frequency appears, is provided by de Haro *et al.* (1996) through numerical experiments for the permeability in porous media. In this paper, we can derive a critical Deborah number $De_c$ in a different way within a parallel plate geometry. Based on the viscoelastic Stokes parameter (3.14) and the accurate bound (3.16), we deduce

$$\Lambda_{ve} = \frac{h}{x_0} < 1 \quad \Leftrightarrow \quad De > \frac{1}{4} \quad (3.18)$$

for any $\bar{\omega}$ in the case of Maxwell fluids. Thus the critical Deborah number can be defined as $De_c = 1/4$. For small $De$ ($De < De_c$), the damping length is small relative to the half height of the channel, and conventional viscous effect dominates. While beyond this critical value, the damping length is large relative to the half height, and the fluid system exhibits viscoelastic behavior. These two flow regimes correspond to 'wide' and 'narrow' systems, respectively.

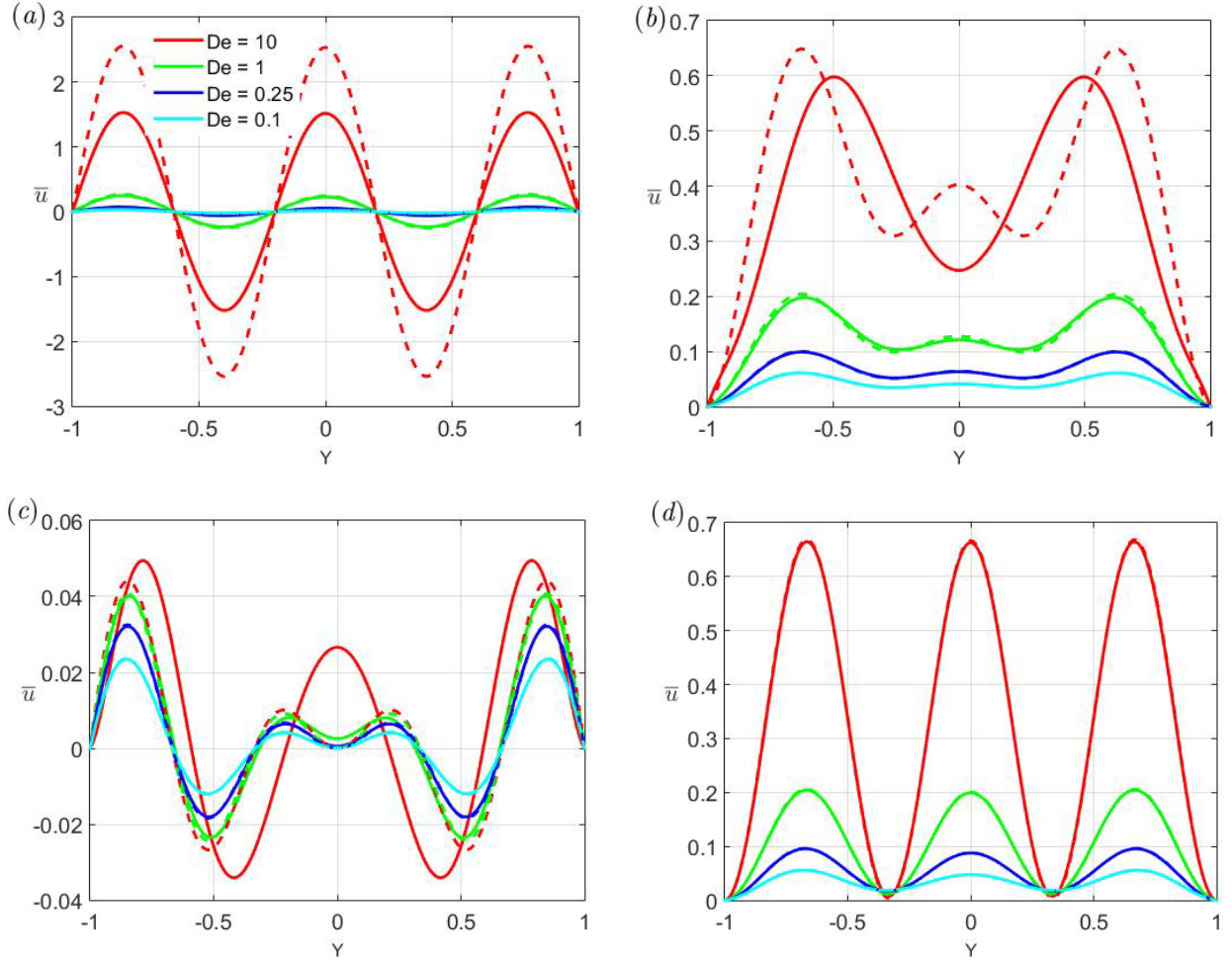

FIGURE 5. (Colour online) Normalized axial velocity profiles of Maxwell fluids ($X = 0$) across the channel at different $De$ number, with the streaming potential effect (solid line) and without the streaming potential effect (dashed line) for: (a) $2h/\lambda_0 = 5/2$, $\omega t = 0$; (b) $2h/\lambda_0 = 5/2$, $\omega t = \pi/2$; (c) $2h/\lambda_0 = 3$, $\omega t = 0$; and (d) $2h/\lambda_0 = 3$, $\omega t = \pi/2$.

To gain some insight into the importance of the above analysis, we present some numerical calculations in the following illustrations, where the typical values of the paramters $K = 15$ and $R = 5$ have been used for microchannels (see (2.25) for the definition of $R$). The profiles of normalized axial velocity $\bar{u}$ are showed in figure 5 at different Deborah number for Maxwell fluids, with and without the streaming potential effect. In panels (a) and (b), we show the velocity profiles satisfying the third-order resonance condition (i.e. $2h/\lambda_0 = 5/2$), at two different time phases $\omega t = 0$ and $\omega t = \pi/2$, respectively. Panels (c) and (d) correspond to the destructive interference condition $2h/\lambda_0 = 3$ at the same two time phases $\omega t = 0$ and $\omega t = \pi/2$, respectively. It can be found that for $De > De_c$ the elastic effect manifests and the oscillating profile is maintained from plate to plate, as a result of the large damping length. For the larger $De$, the elastic effect is more significant. Furthermore, when the constructive interference condition (3.10) is satisfied at the same time, the resonance occurs and the amplitude of the velocity is greatly amplified. At the resonances the velocity is in phase with the harmonic pressure gradient, i.e. the velocity is maximum at $\omega t = 0$ and minimum at $\omega t = \pi/2$ (panels a and b). Conversely, out of the resonances (panels c and d) the velocity is nearly in quadrature with the pressure gradient, and the behavior is reversed, i.e. the velocity is maximum at $\omega t = \pi/2$ and minimum at $\omega t = 0$. In addition, at resonances, the streaming potential effect tends to decrease the magnitude of velocity, especially for large $De$, relative to the case of no streaming potential effect. For example, the velocity is reduced by about 40% at $De = 10$ and $\omega t = 0$ (see figure 5a).

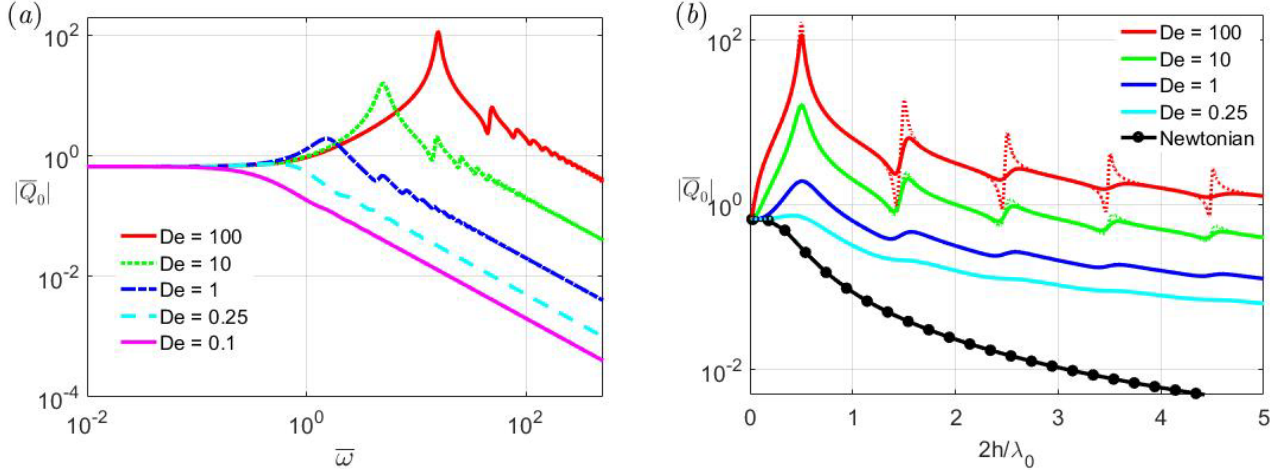

FIGURE 6. (Colour online) Variations of the non-dimensional flow rate of Maxwell fluids at different $De$ number, where two different variables are used as the horizontal axis: (a) non-dimensional frequency $\bar{\omega}$; (b) $2h/\lambda_0$, with the streaming potential effect (solid line) and without the streaming potential effect (dashed line). For comparison, the flow rate for Newtonian fluids ($De = 0$) is also plotted in panel (b).

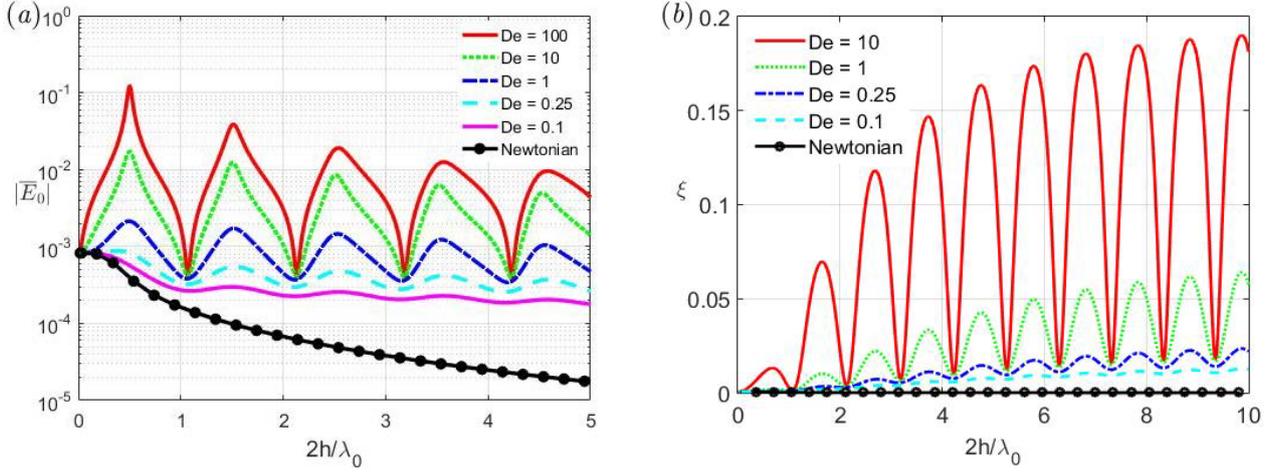

FIGURE 7. (Colour online) Resonance behaviors of (a) non-dimensional streaming potential and (b) EKEC efficiency $\xi$ with the ratio $2h/\lambda_0$ for Maxwell fluids at different $De$ number.

Figure 6 displays the variations of the amplitude of normalized flow rate at different Deborah number, where two different variables are used as the horizontal axis in panels (a) and (b), respectively. It is clear that the appearance of elastic resonance behavior depends on the Deborah number. When $De > De_c$, the resonance may occur at some special frequencies or the values of $2h/\lambda_0$ (see (3.10)). While for $De < De_c$ the resonance does not occur and the flow rate amplitude drops rapidly to zero. As noted by Casanellas & Ortín (2011), the advantage of the choice of $2h/\lambda_0$ instead of the non-dimensional frequency $\bar{\omega}$, for the horizontal axis, is to make the location of the resonant peaks universal (independent of fluid parameters and setup dimensions). The increased frequency $\bar{\omega}$ corresponds to the reduced wavelength $\lambda_0$ and thus the increased ratio $2h/\lambda_0$. In the following illustrations, we will adopt the ratio $2h/\lambda_0$ instead of the frequency as horizontal axis unless otherwise specified. It can be observed that the resonance greatly enhances the flow rate (several orders of magnitude higher than that of Newtonian fluids). Furthermore, the occurrence of resonance and its locations can be characterized well by the conditions given in (3.18) and (3.10), respectively, regardless of whether the streaming potential effect is included (figure 6*b*). Additionally, the streaming potential effect tends to suppress the resonance behavior, and makes the flow rate vary smoothly at resonances, especially for large *De* number.

In figure 7(*a*) we show the variations of the magnitude of the normalized streaming potential as a function of $2h/\lambda_0$ at different *De*. The resonance behavior of the streaming potential can be observed, and it is also determined by the conditions (3.10) and (3.18). At resonances, the amplitude of the streaming potential is amplified greatly, and it is further enhanced as the value of *De* increases. This explains why the influence of the streaming potential is more pronounced at resonances in figure 6(*b*). The variations of the EKEC efficiency $\xi$ for Maxwell fluids as a function of $2h/\lambda_0$ are displayed in figure 7(*b*). It can be found that the viscoelastic behavior of fluids improves the EKEC efficiency. Especially the resonance further amplifies the efficiency by several orders of magnitude, compared with the case of Newtonian fluid. The maximum efficiency is up to 19% in the range of the parameters in our situation. And the resonance behavior of $\xi$ with respect to the forcing frequency of a pressure-driven flow is characterized well by the critical Deborah number $De_c$ and the ratio $2h/\lambda_0$. From figures 6 and 7, it can be observed that as the frequency (or equivalently the ratio $2h/\lambda_0$) grows, the peak values of the flow rate and streaming potential decrease. The maximum of peak values appears at the first-order resonance. However, this is not true for the EKEC efficiency. As the frequency increases, the peak value of the EKEC efficiency gradually increases and eventually tends to saturation (see figure 7*b*).

In figures 5-7, we have shown the elastic resonance phenomena in the electrokinetic flow of Maxwell fluids ($X = 0$) driven by periodic pressure gradient for different Deborah number. Next, we examine the effect of solvent viscosity by using an Oldroyd-B fluid model with positive viscosity ratio $X$ for a fixed Deborah number in figure 8. Note that the critical Deborah number ($De_c = 1/4$) derived based on the Maxwell fluid model is still applicable to the case of Oldroyd-B fluids, which is similar to the multi-mode situation (see figure 9*b*). As can be seen in figure 8, Newtonian solvent contribution tends to suppress the resonance behaviors of the flow rate and the streaming potential in the electrokinetic flow. This can be interpreted as the result of viscous damping and has been observed in a different context (e.g., Andrienko *et al.* 2000; Casanellas & Ortín 2011).

In particular, the high order resonance phenomena are extremely sensitive to the effect of solvent viscosity. For a small ratio $X$ of Newtonian solvent viscosity to the total viscosity of the solution (e.g., $X = 0.01$), the peak values corresponding to the high order resonances are dramatically reduced and even disappear completely. Note that high order resonances occur at high frequencies for fixed channel dimensions and fluid properties. This shows that the solvent viscosity effect is more important in the high frequency regime. In addition, we find that the relatively large solvent viscosity effect (e.g., $X = 0.5$) results in the shift of the first-order resonance position to a lower frequency, relative to the case of $X = 0$ given by the condition (3.10). Furthermore, the EKEC efficiency is dramatically reduced due to the solvent viscosity effect. For example, the maximum efficiency drops rapidly from 19% to 6% when *X* increases from 0 to 0.01 for fixed $De = 10$ (figure 8*c*). In electrokinetic flows of actual viscoelastic fluids (e.g., polymer solutions), the effect of Newtonian solvent is inevitable (refer to Yesilata *et al.* 2006). This means that in practice it is almost impossible to achieve the high efficiency predicted by Bandopadhyay & Chakraborty (2012) and Nguyen *et al.* (2017), although utilizing the resonance behavior of viscoelastic fluids can improve the energy conversion efficiency to some extent.

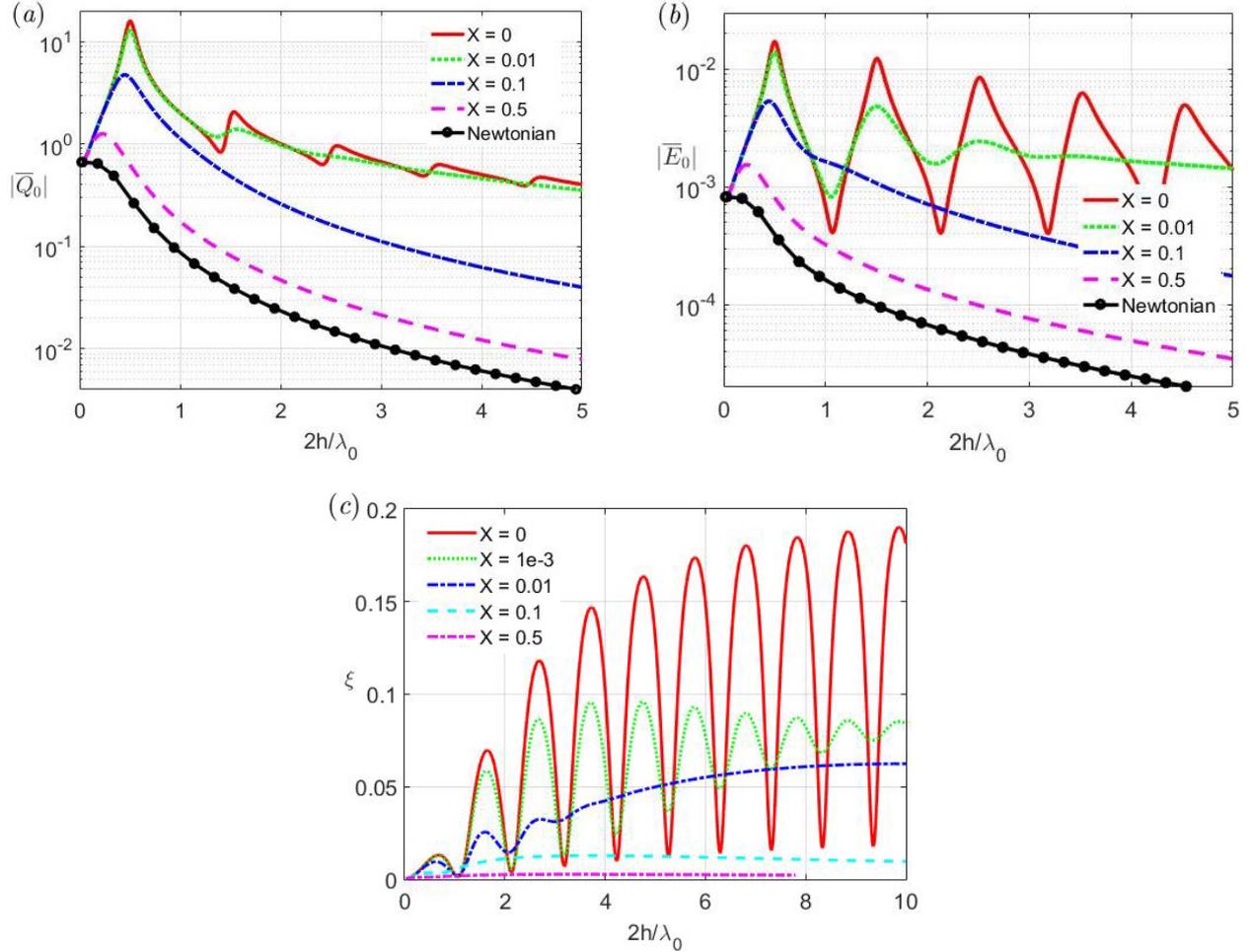

FIGURE 8. (Colour online) Variations of (*a*) the non-dimensional flow rate (*b*) the non-dimensional streaming potential and (*c*) EKEC efficiency $\xi$, vs. the ratio $2h/\lambda_0$ for Oldroyd-B fluid ($De = 10$). Different lines correspond to different values of the viscosity ratio *X*.

### 3.2. *Multi-mode viscoelastic fluids case*

As explained in the introduction, the use of a multi-mode model allows a more realistic description of fluid memory by a spectrum of relaxation times. We now turn our attention to the electrokinetic transport of multi-mode viscoelastic fluids with the constitutive equation (1.1). This model contains constants $\lambda_k$ and $\eta_k$ ($k = 1, \ldots, N$). We adopt the convention that $\lambda_1 > \lambda_2 > \lambda_3 \ldots$ and to reduce the total number of parameters we use the following expressions for generating the relaxation time and viscosity of each mode (Bird *et al.* 1987):

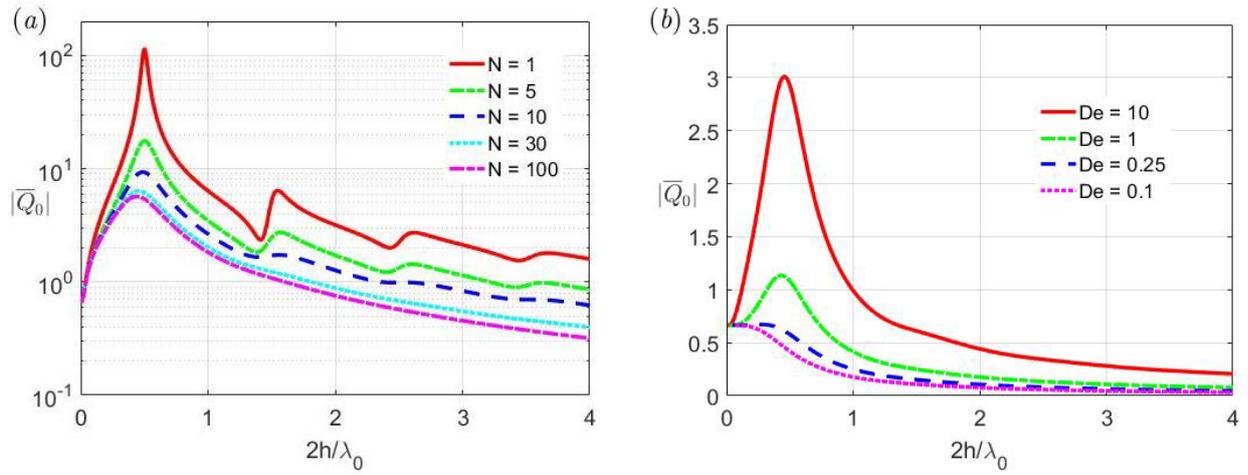

FIGURE 9. (Colour online) Variations of the non-dimensional flow rate for (*a*) different *N* (*De* = 100) and (*b*) different *De* number (*N* = 10). Here the solvent viscosity effect is excluded (*X* = 0).

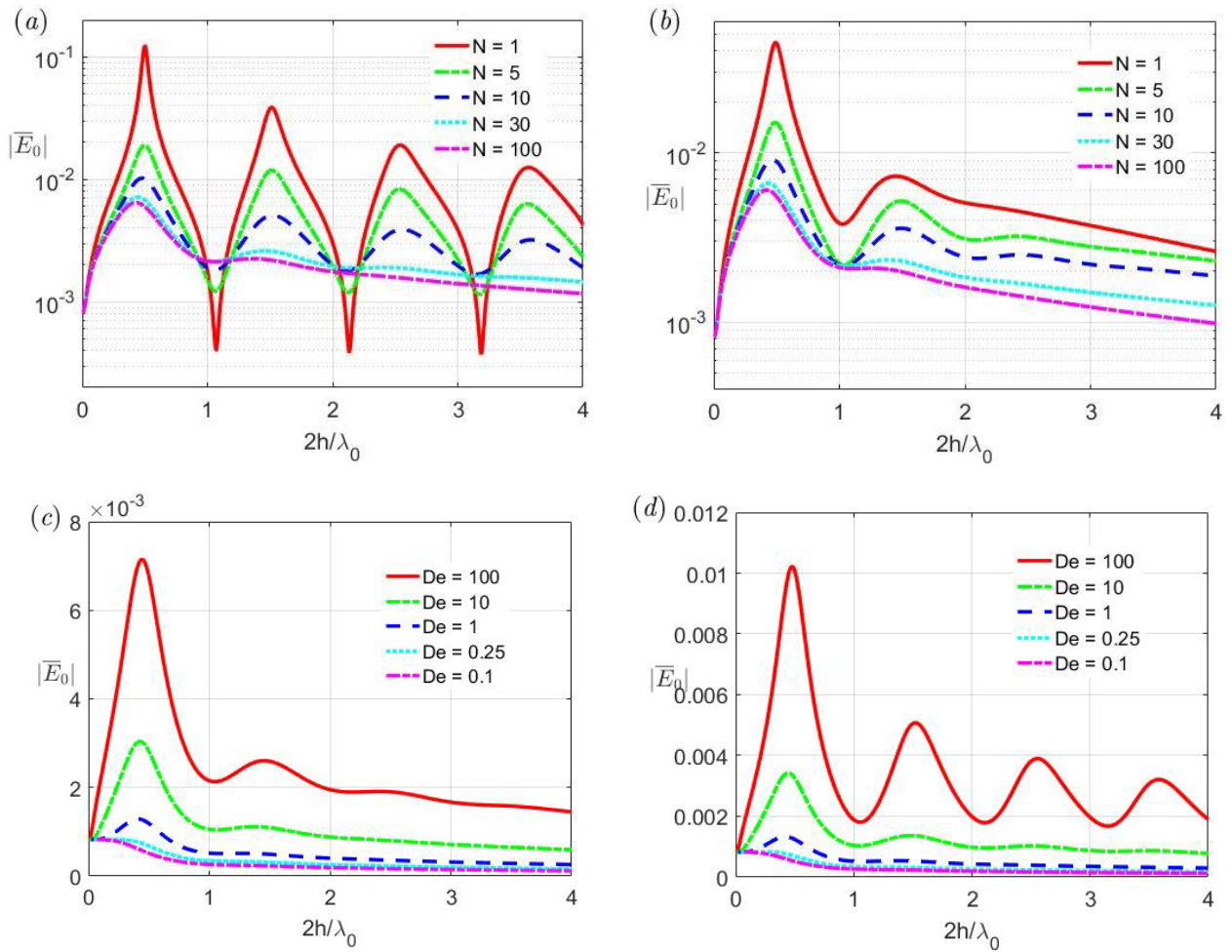

FIGURE 10. (Colour online) Variations of the non-dimensional streaming potential with the ratio $2h/\lambda_0$ for different paramters. Here (*a*) *De* = 100, X = 0; (*b*) *De* = 100, X = 0.01; (*c*) *N* = 30, X = 0; and (*d*) *N* = 10, X = 0.

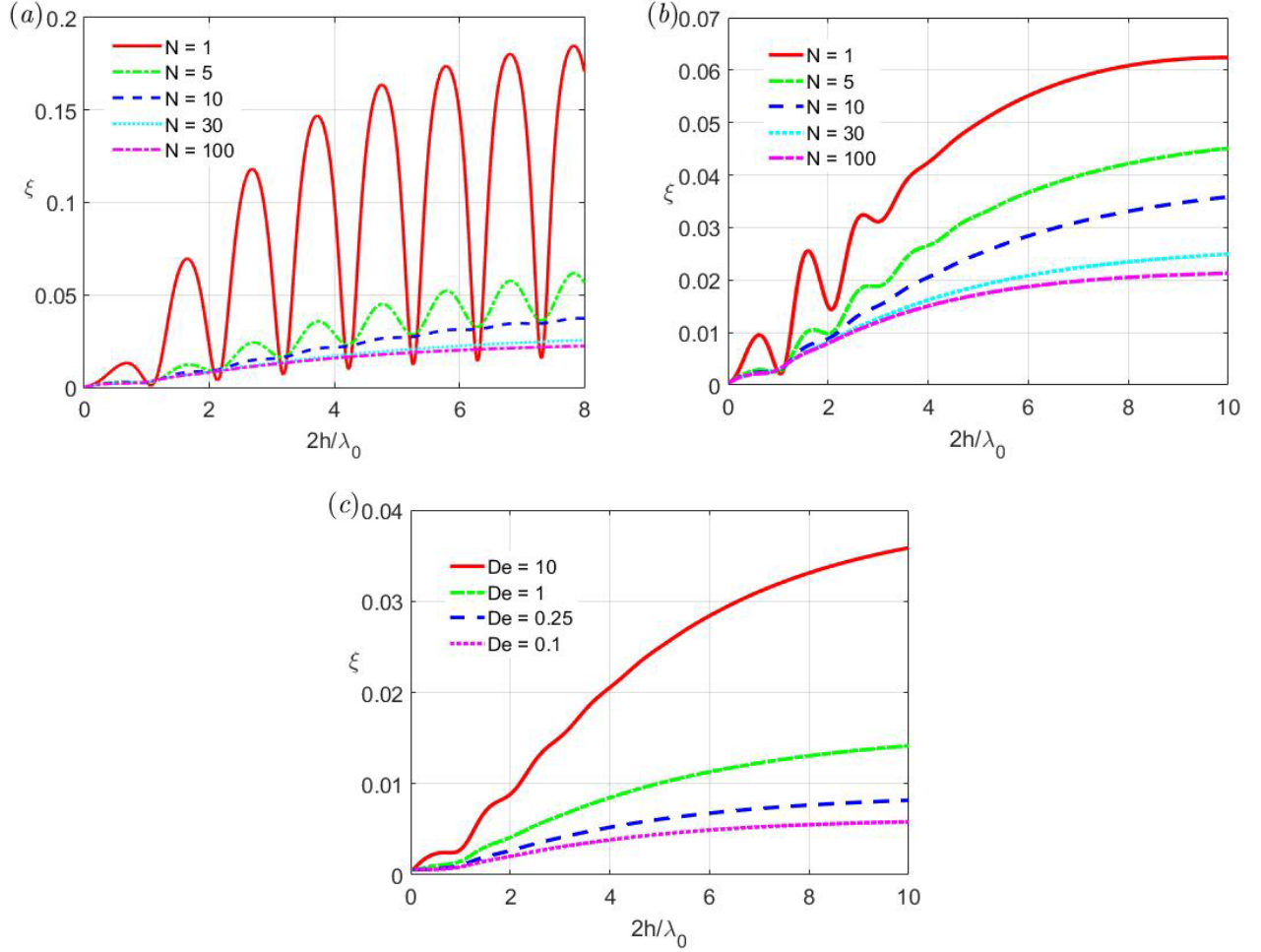

FIGURE 11. (Colour online) EKEC efficiency $\xi$ as a function of $2h/\lambda_0$ for multi-mode viscoelastic fluids at different values of $N$, $X$, and $De$. Here (a) $De = 10$, $X = 0$; (b) $De = 10$, $X = 0.01$; and (c) $N = 10$, $X = 0.01$.

$$\lambda_k = \frac{\lambda}{k^\beta}, \quad \eta_k = \eta_0 \frac{\lambda_k}{\sum_k \lambda_k}, \qquad (3.19)$$

where $\eta_0$ is the zero-shear-rate viscosity, $\lambda$ is a time constant, and $\beta$ is a dimensionless quantity. In all of our computations we use $\beta = 2$ which approximates the value obtained from the Rouse molecular theory for dilute polymer solutions (Prost-Domasky & Khomami 1996). Note that in this model $\eta_p = \sum \eta_k = \eta_0$; $\lambda_k$ decreases as $k$ increases, and the largest relaxation time $\lambda_{max} = \lambda$.

Figure 9(a) illustrates the variations of magnitude of the non-dimensional flow rate for various $N$ values, where $2h/\lambda_0$ has been used for the horizontal axis. We notice that the peak values of resonances decreases with the increase of $N$. This can be attributed to the emergence of a relatively smaller relaxation time, and a role similar to the Newtonian solvent component is played by the smallest relaxation time for large $N$. As $N$ increases, the peak values of high order resonances are dramatically reduced and even disappeared, while the first-order resonance behavior can still be observed for $De > De_c$ although its peak value also decreases. However, unlike the solvent viscosity effect that causes deviation of the resonance locations, in multi-mode case the locations of resonances is well characterized by condition (3.10) and no significant deviation occurs (see also

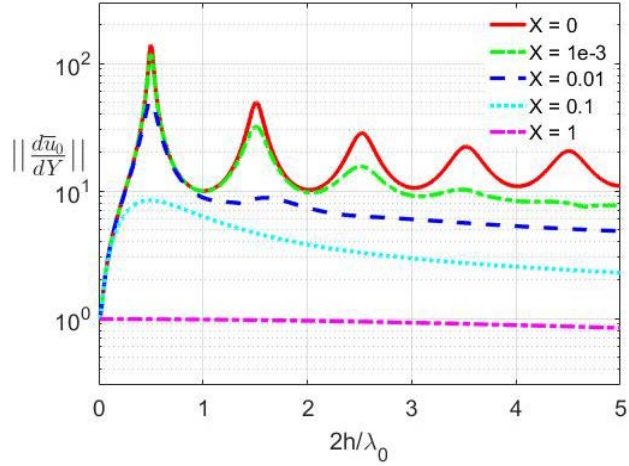

FIGURE 12. (Colour online) Variations of $\|d\bar{u}_0/dY\|$ as a function of $2h/\lambda_0$ for single-mode viscoelastic fluids at different values of $X$, with $De = 100$. Note that $X = 1$ corresponds to Newtonian fluids case.

figure 10). Dependence of the resonance behavior of flow rate on the Deborah number is displayed in figure 9(*b*). It can be found that the critical Deborah number $De_c = 1/4$ derived from the single-mode Maxwell fluid model is still applicable to the multi-mode case. When $De > De_c$, the resonance occurs, while for $De < De_c$ the resonance disappears and the flow rate amplitude drops rapidly to zero with the frequency.

The variations of the non-dimensional streaming potential magnitude and the EKEC efficiency as a function of $2h/\lambda_0$ at different parameters $N$, $De$ and $X$ are demonstrated in figures 10 and 11 respectively. Multiple relaxation times suppress the elastic resonance behaviors of the streaming potential and EKEC efficiency, compared with the single-mode Maxwell fluids. Similar to the flow rate, the occurrence of resonance for the streaming potential field is dominated by the critical Deborah number. When $De < De_c$ the resonance disappears and the amplitude of streaming potential field drops to zero with the frequency (see figures 10*c* and 10*d*). This can be explained as follows: the largest one in the relaxation time spectrum determines the elastic nature of the fluid, so the Deborah number defined based on the maximum relaxation time can capture the appearance of elastic resonance phenomena well. However in the multi-mode case, the occurrence of resonance for the efficiency is complex. It is dominated by the Deborah number $De$ and the mode number $N$. For small $De$ or large $N$ (e.g., $N > 10$ in figure 11), the resonance behavior of efficiency will disappear.

## 4. Validity of the linear analysis

In this paper, we investigate the electrokinetic phenomena of viscoelastic fluids driven by periodic pressure gradient based on multi-mode models, which include the single-mode Maxwell and Oldroyd-B constitutive equations as particular cases. We have assumed that the fluid is in the laminar oscillatory flow so that the governing equation of fluid velocity is reduced to the linear one and the fields of velocity and streaming potential can be solved analytically. In general, the instability and transition of viscoelastic flows may occur in large amplitude oscillatory shear, which depends on the geometry of the channel, $Wi$ and $Re$. In micro-scale flows, $Re$ is usually small and can be ignored, so it is the Weissenberg number $Wi$ that plays an important role in this basic geometry. Meanwhile, a critical Weissenberg number $Wi_c$ related to the stability can be introduced.

| $De$ | $X$ | $\|\mathrm{d}\bar{u}_0/\mathrm{d}Y\|$ | $Y_1$ | $\Delta p^*$ (bar) |
|---|---|---|---|---|
| 100 | 0 | 139.28 | 1 | 0.14 |
|  | 0.001 | 118.58 | 0.95 | 0.17 |
|  | 0.01 | 50.94 | 0.92 | 0.39 |
|  | 0.1 | 8.45 | 0.97 | 2.37 |
|  | 0.9 | 1.09 | 1 | 18.30 |
| 10 | 0 | 19.45 | 0.96 | 1.03 |
|  | 0.001 | 19.00 | 0.96 | 1.05 |
|  | 0.01 | 15.76 | 0.96 | 1.27 |
|  | 0.1 | 6.10 | 0.98 | 3.28 |
|  | 0.9 | 1.07 | 1 | 18.68 |

TABLE 1. Maximum values of the dimensionless velocity gradient at the first-order resonance for different $De$ and $X$. Here $Y_1$ represents the position of the maximum value across the channel, and $\Delta p^*$ represents a critical pressure difference applied to the system, where $Wi_c = 1$ has been used. Note that for $X = 0.9$ it is close to Newtonian fluid case.

Typically, $Wi_c \sim O(1)$ (Torralba et al. 2007). For $Wi < Wi_c$ the fluid is expected to be in parallel shear flows with straight streamlines, while above this threshold nonlinearities and elastic instabilities start to become manifest. In this section, we examine the validity of our linear analysis based on this threshold Weissenberg number, which places a limit on the maximum pressure gradient (or pressure difference) applied to the fluid system as showed later.

According to the definition, we have

$$Wi = \lambda_{\max}\dot{\gamma}_{\max} = \lambda_{\max}\max\left\{\left|\frac{\partial u}{\partial y}\right|\right\} = \frac{\lambda_{\max}h}{\eta}\frac{\mathrm{d}p_0}{\mathrm{d}x}\max\left\{\left|\frac{\mathrm{d}\bar{u}_0}{\mathrm{d}Y}\right|\right\}, \quad (4.1)$$

where the local shear rate is computed by the velocity gradient and the dimensionless variables have been used. The dimensionless velocity gradient is readily available by differentiating the expression (2.17) of velocity. For simplicity, we first consider the single-mode fluid case, i.e., $\lambda_{\max} = \lambda$. Furthermore, using the Cox-Merz rule, $\eta \simeq G_0\lambda$, an empirical relation that allows to relating complex properties obtained under oscillatory shear experiments with steady shear flow measurements at corresponding values of frequency and shear rate (Bird et al. 1987), we get

$$Wi = \frac{h}{G_0}\frac{\mathrm{d}p_0}{\mathrm{d}x}\left\|\frac{\mathrm{d}\bar{u}_0}{\mathrm{d}Y}\right\|, \quad (4.2)$$

with the norm $\|\mathrm{d}\bar{u}_0/\mathrm{d}Y\| = \max\{|\mathrm{d}\bar{u}_0(Y)/\mathrm{d}Y|\}$. Figure 12 illustrates the variations of $\|\mathrm{d}\bar{u}_0/\mathrm{d}Y\|$ as a function of $2h/\lambda_0$ for different values of $X$. Undoubtedly, peak values of the velocity gradient (or shear rate) are experienced at resonances, and the largest one is at the first-order resonance. Furthermore, the largest values corresponding to different parameters are given in table 1. The constant $G_0$ depends on rheological properties of fluids, and its value varies from 0.45 Pa to 30.2 Pa according to small-amplitude oscillatory shear experiments of the wormlike micellar fluid samples (see Yesilata *et al*. 2006).

In order to make the fluid response in the linear regime, the Weissenberg number of the flow is required to be less than the corresponding critical value, as stated above. This implies that the magnitude of the imposed pressure gradient must satisfy

$$\frac{\mathrm{d}p_0}{\mathrm{d}x} < \frac{G_0 Wi_c}{h} \left\| \frac{\mathrm{d}\bar{u}_0}{\mathrm{d}Y} \right\|^{-1}. \tag{4.3}$$

Accordingly, we can define a critical pressure difference through the channel as $\Delta p^* = lG_0 Wi_c \|\mathrm{d}\bar{u}_0/\mathrm{d}Y\|^{-1}$ with the length $l$ of the channel. For illustrative computations, we consider the length ($l$) and height ($h$) of the channel are 100 mm and 1 $\mu$m, respectively, and $G_0 = 20$ Pa. The results are shown in table 1. It can be found that the value of $\|\mathrm{d}\bar{u}_0/\mathrm{d}Y\|$ increases with $De$ and decreases with $X$, which is consistent with the effects of parameters $De$ and $X$ on the viscoelastic behavior. The maximum velocity gradient occurs near the walls (table 1). Moreover, the critical pressure difference is small when the viscoelastic effect of fluids is significant (e.g., large $De$ and tiny $X$). On the other hand, when the fluid behavior is close to Newtonian fluid case, the critical pressure difference increases significantly.

Finally, we extend to the case of multi-mode viscoelastic fluids containing a spectrum of relaxation times ($N > 1$). Since the viscoelastic behavior is weakened compared to the single-mode fluid with the largest relaxation time ($N = 1$), peak values of the velocity gradient or shear rate at resonances are reduced. Therefore the critical pressure difference, determining whether the fluid is in the laminar oscillatory flow, is increased compared to the single-mode case.

## 5. Summary and conclusions

In this study, we investigate the resonance phenomenon in electrokinetic transport of viscoelastic fluids that include the realistic effects of Newtonian solvent and multiple relaxation times. Under the assumptions we considered (e.g., low EDL potential and small $Wi$), the governing equations of fluid velocity and electric potential are reduced to a set of linear equations, which can be solved analytically. In the analytical solutions, we introduce a complex dimensionless parameter $\alpha$ that plays a key role in the analysis. By means of the equivalence between the time-periodic pressure driven flow and the flow induced by the synchronous oscillatory sidewalls in the linear regime, the meaning of $\alpha$ can be explained from the perspective of viscoelastic shear waves generated by the oscillatory sidewalls. Its real part represents the ratio of the height of the channel to the wavelength of the shear waves, and the imaginary part represents the ratio of the half height of the channel to the damping length of the shear waves. Based on the imaginary part of $\alpha$ one can define a viscoelastic Stokes parameter that determines whether the interference of waves generated at the opposite walls can occur. If the interference occurs, the interference form (i.e. constructive or destructive interference) depends on the real part of $\alpha$. When the conditions of constructive interference are satisfied, the resonance appears. Furthermore, the following conclusions can be drawn.

First, based on the interaction of viscoelastic shear waves, we explain the mechanism of the resonance, and derive a critical Deborah number $De_c = 1/4$ which dominates the occurrence of resonance and is universal for parallel plate geometries regardless of whether the streaming potential effect is included. The streaming potential effect does not change the 'locations' of resonances, but will weaken the strength of resonances. Above this threshold ($De_c$) the resonances appear at particular frequencies and result in a dramatic amplification of the amplitudes of flow rate and streaming potential. Particularly the resonance further amplifies the EKEC efficiency by several orders of magnitude, compared with the case of Newtonian fluid. The maximum efficiency is up to 19% in the range of the parameters in our situation. In addition, the concept of resonance order has been introduced to accurately describe the behavior of resonances corresponding to the different driving frequencies. The locations of resonant peaks with different order are determined universally by the ratio $2h/\lambda_0$.

Second, the effect of solvent viscosity is examined by using the Oldroyd-B fluid model. The Newtonian solvent contribution tends to suppress the resonance behaviors of the flow rate, streaming potential and EKEC efficiency in the electrokinetic flow. This can be interpreted as a result of viscous damping. For example, the maximum efficiency drops rapidly from 19% to 6% for extremely small ratios of the solvent to the solution viscosity ($X = 0.01$). This means that the solvent viscosity effect greatly hinders the realization of high efficiency predicted by Bandopadhyay & Chakraborty (2012) and Nguyen *et al*. (2017). In addition, the high order resonance phenomena are extremely sensitive to the effect of solvent viscosity. For a small ratio *X*, the peak values corresponding to the high order resonance are dramatically reduced and even disappeared.

Third, we investigate the streaming potential and EKEC of viscoelastic fluids by using a multi-mode model that allows for a spectrum of relaxation times and viscosities. Multiple relaxation times show inhibitory effect on the elastic resonance behaviors for the streaming potential field and EKEC efficiency, compared with the single-mode one. We notice that their peak values at resonances decreases with the increase of the number of relaxation times in the spectrum, which corresponds to the emergence of a relatively smaller relaxation time. Furthermore for the streaming potential field and flow rate, the occurrence of resonance behaviors are still dominated by the critical Deborah number $De_c$. The locations of resonances are also well characterized by the ratio $2h/\lambda_0$. However in the multi-mode case, the occurrence of resonance behavior for the EKEC efficiency is complex. It is dominated by the Deborah number *De* and the mode number *N*. For small *De* or large *N*, the resonance behavior of efficiency will disappear completely.

In addition, we have identified three distinct regimes for the scaling relations between the dimensionless streaming potential field $\bar{E}_{0,\max}$ and EDL thickness $K^{-1}$ at high Deborah number: when the EDL thickness is much less than the half height of the channel ($K^{-1} < 0.1$), $\bar{E}_{0,\max}$ exhibits a quadratic growth trend with respect to $K^{-1}$ (Quadratic Regime); when the EDL thickness is comparable to the half height of the channel (i.e. $K^{-1} \approx 1$), $\bar{E}_{0,\max}$ increases linearly with $K^{-1}$ (Linear Regime); for larger overlapped values of the EDL thickness ($K^{-1} > 10$), $\bar{E}_{0,\max}$ saturates and does not vary with $K^{-1}$ (Saturation Regime).

Finally, we check the validity of the linear analysis, which requires small driving amplitudes. Based on a threshold Weissenberg number, the maximum pressure gradient (or pressure difference) applied to fluid systems can be estimated. If the driving amplitudes exceed this threshold, nonlinear rheology (e.g., shear-thinning behavior) becomes relevant, which deserves to be further studied. Our future work will focus on the nonlinear features of elastic resonances in the electrokinetic flow of viscoelastic fluids.

**Acknowledgements**

We acknowledge financial support from the National Natural Science Foundation of China (Grant Nos. 11902165, 11772162), the Natural Science Foundation of Inner Mongolia Autonomous Region of China (Grant No. 2019BS01004) and the Inner Mongolia Grassland Talent (Grant No. 12000-12102408).

**Declaration of interests**

The authors report no conflict of interest.

**Appendix A. The equivalence between the time-periodic pressure driven electrokinetic flow and the flow induced by the synchronous oscillatory sidewalls**

The governing equations of fluid flow driven by periodic pressure gradient are given by

$$\left.\begin{array}{r}\rho\dfrac{\partial \boldsymbol{V}}{\partial t}+\rho(\boldsymbol{V}\cdot\nabla)\boldsymbol{V}=-\nabla p+\nabla\cdot\boldsymbol{T}+\boldsymbol{F}_{EK},\\ \boldsymbol{T}=\boldsymbol{T}_s+\boldsymbol{T}_p,\\ \boldsymbol{T}_s=\eta_s\dot{\boldsymbol{\gamma}},\ \boldsymbol{T}_p=\sum_{k=1}^{N}\boldsymbol{T}_k,\\ \boldsymbol{T}_k+\lambda_k\stackrel{\nabla}{\boldsymbol{T}}_k=\eta_k\dot{\boldsymbol{\gamma}},\end{array}\right\} \quad (A1)$$

subjected to the no-slip boundary conditions

$$\boldsymbol{V}\big|_{\text{walls}}=0. \quad (A2)$$

Here $\boldsymbol{F}_{EK}=\rho_e \boldsymbol{E}$ represents electric field force; $\dot{\boldsymbol{\gamma}}=\nabla\boldsymbol{V}+(\nabla\boldsymbol{V})^{\dagger}$ is the rate-of-strain tensor.

Under the assumption of fully developed unidirectional flow, the above equations (A1) and (A2) can be reduced to

$$\left.\begin{array}{r}\rho\dfrac{\partial u}{\partial t}=-\dfrac{\partial p}{\partial x}+\eta_s\dfrac{\partial^2 u}{\partial y^2}+\sum_{k=1}^{N}\dfrac{\partial \tau_k}{\partial y}+\rho_e E_s,\\ \left(1+\lambda_k\dfrac{\partial}{\partial t}\right)\tau_k=\eta_k\dfrac{\partial u}{\partial y},\\ u\big|_{\text{walls}}=0,\end{array}\right\} \quad (A3)$$

where $u(y, t)$ is the axial velocity, and $\partial p/\partial x = A\cos(\omega t)$ is the harmonic pressure gradient independent of position. The nonlinear convection term in (A1) has been vanished, so the problem is in the linear regime.

On the other hand, consider the electrokinetic flow induced by the synchronous oscillatory sidewalls with the same frequency. Under the same assumption of fully developed unidirectional flow, the governing equations can be written as

$$\left.\begin{aligned}\rho\frac{\partial \tilde{u}}{\partial t} &= \eta_s \frac{\partial^2 \tilde{u}}{\partial y^2} + \sum_{k=1}^{N} \frac{\partial \tau_k}{\partial y} + \rho_e E_s, \\ \left(1+\lambda_k \frac{\partial}{\partial t}\right)\tau_k &= \eta_k \frac{\partial \tilde{u}}{\partial y}, \\ \tilde{u}|_{\text{walls}} &= B\sin(\omega t),\end{aligned}\right\} \quad (A4)$$

where *B* is the amplitude.

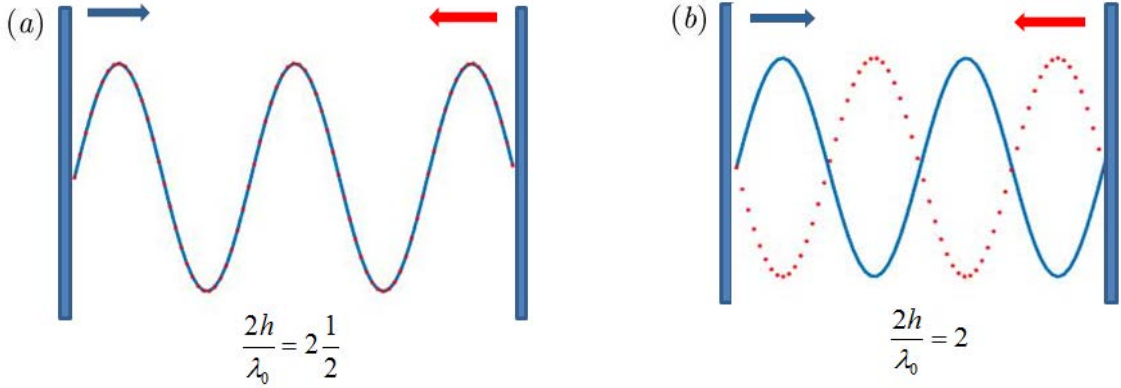

FIGURE 13. (Colour online) Schematic diagram for the interference phenomenon of viscoelastic shear waves generated at the opposite walls. (*a*) represents the constructive interference of the shear waves where the size of the channel 2*h* is a half integer multiple of the wavelength $\lambda_0$; (*b*) represents the destructive interference of the shear waves where the size of the channel is an integer multiple of the wavelength. The arrows indicate the direction of wave propagation.

After introducing the transformation $\tilde{u} = u + B\sin(\omega t)$, one can readily find that the equations in (A4) are the same with the ones in (A3) given the relationship $B = A/\rho\omega$ holds. This implies that the time-periodic pressure driven electrokinetic flow is equivalent to the flow that results from the synchronous oscillatory sidewalls with the same frequency, observed in the reference frame of the sidewalls. Here it is worth noting that the the oscillating reference frame is not an inertial frame of reference. For an incompressible flow in which the density $\rho$ is a constant, the inertial body force term $\rho\omega B\cos(\omega t)$, generated after the transformation $\tilde{u} = u + B\sin(\omega t)$ in the equation (A4), can be regarded as a pressure gradient $\partial p/\partial x = \rho\omega B\cos(\omega t)$, and thus can be included in the Navier-Stokes equations, thereby recovering exactly the same formulation as in (A3) with a fixed sidewall. In addition, the above equivalence does not depend on the geometry of channel section. However, we point out that if beyond the linear regime (e.g., the convection term does not vanish), this equivalence will no longer hold.

Based on this equivalence, the flow can be viewed as the result of the interference in time and space of the shear waves generated by the oscillatory sidewalls. In figure 13, we show the schematic diagram for the interference phenomenon of viscoelastic shear waves generated at the opposite walls. Here, the damping of viscoelastic shear waves is ignored and we focus on the interference behavior between them, especially constructive and destructive interferences.